\documentclass{aa}

\usepackage{graphicx}
\usepackage{txfonts}

\usepackage{mathrsfs}
\usepackage{amsmath}
\usepackage{bm}

\defcitealias{wang2022}{I}

\newcommand{\dotm}{\ifmmode {\dot{\mathscr{M}}} \else $\dot{\mathscr{M}}$\fi}

\begin{document}

\title{Spiral Arms in Broad-line Regions of Active Galactic Nuclei}


\subtitle{II. Loosely Wound Cases: Reverberation Properties}

\author{Pu Du
    \inst{1}
    \and
    Jian-Min Wang
    \inst{1, 2, 3}}

\institute{
    Key Laboratory for Particle Astrophysics,
    Institute of High Energy Physics, Chinese Academy of Sciences,
    19B Yuquan Road, Beijing 100049, China \\
    \and
    University of Chinese Academy of Sciences,
    19A Yuquan Road, Beijing 100049, China \\
    \and
    National Astronomical Observatories of China,
    Chinese Academy of Sciences, 20A Datun Road, Beijing 100020, China}


\abstract{There has recently been growing evidence that broad-line regions (BLRs) 
in active galactic nuclei have regular substructures, such as spiral arms, which 
are supported by the fact that the radii of BLRs measured by RM observations 
are generally consistent with the self-gravitating regions of accretion disks. 
We have shown in Paper I that the spiral arms excited by the gravitational 
instabilities in these regions may exist in some disk-like BLRs. As the second 
paper of the series, we investigate the loosely wound spiral arms excited by 
gravitational instabilities in disk-like BLRs and present their observational 
characteristics. Following the treatments of \cite{adams1989}, we solve the governing 
integro-differential equation by a matrix scheme. The emission-line profiles, 
velocity-delay maps, and velocity-resolved lags of the 
BLR spiral arms are calculated. We find that the spiral arms can explain some 
phenomena in observations: (1) the emission-line profiles in the mean and rms 
spectra have different asymmetries, (2) some velocity-delay 
maps, e.g., NGC 5548, have complex sub-features (incomplete ellipse), (3) the 
timescales of the asymmetry changes in emission-line profiles (rms spectra) are short. 
These features are attractive for modeling the observed line profiles and the
properties of reverberation, and for revealing the details of the BLR geometry and kinematics.}


\maketitle

\section{Introduction}
\label{sec:Intro}

As the prominent features in the UV/optical spectra of active galactic nuclei
(AGNs), the broad emission lines with velocity widths of $\sim$1000 -- 20000 $\rm
km~s^{-1}$ originate from the broad-line regions (BLRs) photoionized
by the continuum radiation from the central accretion disks around supermassive
black holes (SMBHs). The physics of BLRs (e.g., the geometry, kinematics, mass
distributions, and photoionization properties), which
determines the profiles of broad emission lines, is not only related with the
origin and evolution of the materials in the central regions of AGNs, but also
closely connected with the measurement of BH masses in reverberation mapping
\citep[RM, e.g.,][]{blandford1982, peterson1993}. It makes BLRs one of the core
topics in AGN researches. 

RM is a technique to probe the geometry and kinematics of BLRs and to measure
the masses of SMBHs in AGNs. It has been successfully applied to more than 100
objects in the past decades \citep[e.g.,][]{peterson1998, kaspi2000, bentz2009,
denney2009, barth2011, rafter2011, du2014, du2018, fausnaugh2017, grier2017a,
derosa2018, rakshit2019, hu2021, yu2021, bao2022}. RM measures the delayed
response of broad emission lines with respect to the varying continuum emission.
Due to the limits of spectral resolution and flux calibration precision, most RM
campaigns in the early days focused on the average time lags ($\tau_{\rm
H\beta}$) of H$\beta$ emission line \citep[e.g.,][]{peterson1998, kaspi2000}. In
combination with the velocity widths ($V_{\rm H\beta}$) of H$\beta$ lines, the
masses of SMBHs can be formulated with $M_{\bullet} = f_{\rm BLR} V_{\rm
H\beta}^2 R_{\rm H\beta} / G$, where $R_{\rm H\beta} = c \tau_{\rm H\beta}$ is
the emissivity-weighted radius of BLR, $c$ is the speed of light, $G$ is the
gravitational constant, and $f_{\rm BLR}$ is a parameter called ``virial
factor'', which is controlled by the BLR geometry and kinematics. Therefore, the
accuracy of BH mass measurement is directly related with the understanding of
BLR physics. Furthermore, with the improvement of flux calibration and spectral
resolution in recent years, velocity-resolved RM, rather than only measuring an
average time lag, has been gradually performed to more and more objects. It aims
to measure the time lag as a function of velocity \citep[e.g.,][]{bentz2008,
bentz2010a, denney2010, du2016VI, du2018b, pei2017, derosa2018, hu2020a,
hu2020b, brotherton2020, lu2021, u2022, bao2022} or, more importantly, to reconstruct
the ``velocity-delay maps'' (also known as transfer functions) of BLRs by
model-independent methods such as maximum entropy method
\citep[e.g.,][]{bentz2010b, grier2013, skielboe2015, xiao2018a, xiao2018b,
brotherton2020, horne2021} or to constrain the BLR parameters by the Bayesian
modeling through Markov Chain Monte Carlo \citep[MCMC, e.g.,][]{pancoast2012,
pancoast2014, grier2017b, williams2018, li2018, villafana2022}. The general
geometry and kinematics of the BLRs (e.g., disk-like, inflow, or outflow) in
dozens of AGNs have been successfully revealed (see the aforementioned
references).

Systematic researches on the inhomogeneity and sub-structures in BLRs are
relatively scarce, however, their signs in observations have appeared
gradually. Three pieces of evidence imply the existence of the
inhomogeneity and sub-structures in BLRs. (1) Many AGNs show complex
emission-line profiles, even with multiple wiggles or small peaks, rather than
symmetric profiles or simply asymmetric profiles with a little stronger red or
blue wings in their emission lines (e.g., the line profiles of Mrk 6, Mrk 715,
or NGC 2617 in the Appendix of \citealt{du2018b}). It indicates that the BLR gas
distributions in those objects should be more complex than previously thought.
(2) There is a well-known phenomenon that the line profiles in the mean and rms
spectra of RM are commonly different for a same object
\citep[e.g.,][]{peterson1998, bentz2009, denney2009, barth2013, fausnaugh2017,
grier2012, du2018b, derosa2018, brotherton2020}. The profiles of the emission
lines in rms spectra represent the geometry and kinematics of the gas that has
response to the continuum variations and only is a portion of the total BLR
gas. The differences between the mean and rms spectra suggest the gas
inhomogeneity in BLRs. (3) More importantly, the velocity-delay maps of some
objects (e.g., NGC 5548) have shown complex features (e.g., incomplete ellipse,
bright strips) in comparison with the simple disk, inflow, or outflow models.
They are probably the evidence of the BLR inhomogeneity and sub-structures
\citep[e.g.,][]{xiao2018b, horne2021}.

The radii of BLRs measured by RM mostly span from $10^3 R_{\rm g}$ to $10^5
R_{\rm g}$ for different objects, where $R_{\rm g} = 1.5 \times 10^{13} M_8 \
{\rm cm}$ is the gravitational radius and $M_8 = M_{\bullet} / 10^8 M_{\odot}$
is the SMBH mass in unit of $10^8$ solar mass \citep{du2016V}.  Such a range of
radius is consistent with the self-gravitating region of accretion disk
\citep[e.g.,][]{paczynski1978, shlosman1987, bertin1999, goodman2003,
sirko2003}. Besides, a number of objects (e.g., Arp 151, 3C 120, NGC 5548) show
clear RM signatures of Keplerian disks \citep[][]{bentz2010b, grier2013,
xiao2018b, horne2021}. The heuristic idea that the origin of BLRs is related
with the self-gravitating regions of accretion disks was initially discussed by
\cite{shore1982}, and was further theoretically studied in the subsequent works
\citep[e.g.,][]{collin1987, collin1990, dumont1990a, dumont1990b}.  Although the
detailed physics in the self-gravitating region is still far from fully
understood, the existence of spiral arms may be a natural consequence resulted from
the gravitational instabilities in this region \citep[e.g.,][]{lodato2007}. 

On the other hand, the mass ratio of standard accretion disk \citep{shakura1973}
to SMBH can be expressed as $0.04 \alpha_{0.1}^{-4/5} M_8^{6/5} \dotm^{7/10}
r_4^{5/4}$ (or $0.7 \alpha_{0.1}^{-4/5} M_8^{6/5} \dotm^{7/10} r_5^{5/4}$
depending on the typical radius) if the disk extends to the scale size of BLR,
where $\dotm = \dot{M}_{\bullet} c^2 / L_{\rm Edd}$ is the dimensionless
accretion rate, $\dot{M}_{\bullet}$ is the mass accretion rate, $L_{\rm Edd} =
1.5 \times 10^{46} M_8 \ {\rm erg\ s^{-1}}$ is the Eddington luminosity of solar
composition gas, $\alpha_{0.1} = \alpha / 0.1$ is the viscosity parameter, and
$r_4 = R_{\rm out} / 10^4 R_g$ (or $r_5 = R_{\rm out} / 10^5 R_g$) is the outer
radius. This ratio is generally similar to the disk-to-star mass ratios in
protoplanetary systems, which commonly possess spiral arm structures
\citep[e.g.,][]{andrews2013, dong2018}. This also leads to the possibility that
BLRs can host spiral arms. 

Therefore, it is important to investigate the spiral arms in BLRs and their
potential characteristics in observations. \cite{horne2004} calculated the
velocity-delay map of a photoionized disk with two spiral arms mathematically
without introducing any precise physics (through ``twisting'' the elliptical
orbits). \cite{gilbert1999}, \cite{storchi-bergmann2003}, \cite{schimoia2012},
and \cite{storchi-bergmann2017} assumed an analytical form of the spiral arms
and explained the double-peaked profiles of the broad emission lines in AGNs,
but similarly do not include any dynamical physics. As the first paper of
this series, \cite{wang2022} introduced the density wave theory of spiral
galaxies \citep[e.g.,][]{lin1964, lin1966, lin1969}, which applies to
self-gravitating disks \citep{goldreich1979}, into the research of BLRs for the
first time (hereafter Paper \citetalias{wang2022}). Paper
\citetalias{wang2022} explores the possibility of density waves in BLRs
through discussing their physical conditions, and focuses on the simplest cases of
tight-winding arms with short wavelengths and small pitch angles (adopting the
formalism of tight-winding approximation). However, the loosely wound spiral
arms have more significant features in line profiles or RM signals relative to
the tightly wound cases (see more details in Paper \citetalias{wang2022} or in the
following sections of the present paper). Hence, it is crucial to investigate
the loosely wound spiral arms in BLRs and their characteristics in observations.

As the second paper of this series, here we calculate the surface density
distributions of loosely wound spiral arms in a numerical manner without
the tight-winding approximation, and their corresponding emission-line
profiles, velocity-delay maps, and velocity-resolved lags. Comparing with
Paper \citetalias{wang2022}, we adopt more general radial distributions of the
BLR surface density and sound speed, which are assumed as power laws with free
indexes. This is a natural extension of Paper \citetalias{wang2022}. The paper
is organized as follows. In Section \ref{sec:Theory}, we briefly introduce the
density wave model and the numerical method. Some fiducial modes (arm patterns)
and their observational signals (in emission-line profiles, velocity-delay maps,
and velocity-resolved lags) for different azimuthal angles of the line of sight
(LOS) are provided in Sections \ref{sec:spiral_arms} and
\ref{sec:obs_characteristics}. We discuss and compare the models with the
observations in Section \ref{sec:discussion}. A brief summary is given in
Section \ref{sec:summary}.

\section{Theoretical Formulation}
\label{sec:Theory}

We adopt the density wave formalism in \cite{Lin1979} and the numerical method
in \cite{adams1989} to calculate the spiral arms. The perturbation equations and
numerical method in \cite{adams1989} apply to both tightly and loosely wound
arms, and can also derive one-armed density wave (azimuthal wave number $m=1$).
Details of the formula deduction and numerical procedures can be found in these
papers and the references therein. For completeness, we briefly describe the
key points in this section. The model in the present paper assume the general
geometry of BLR is disk-like. It may apply to the objects which show clear
features of Keplerian disks in their RM signals \citep[e.g., Arp 151, 3C 120,
NGC 5548 in][]{bentz2010b, grier2013, xiao2018b, horne2021}.

\subsection{Perturbation Equations}
\label{sec:perturbation_equations}

Here we adopt the linear normal-mode formalism in
\cite{adams1989} (also refer to the more recent work of \citealt{chen2021}). We
use the cylindrical coordinates ($R$, $\varphi$, $z$). In a thin disk, the
continuity equation and the motion equations in radial and azimuthal directions
read
\begin{equation}
    \frac{\partial\sigma}{\partial t} + \frac{1}{R}\frac{\partial}{\partial R}(R\sigma u) 
    + \frac{1}{R}\frac{\partial}{\partial\varphi}(\sigma \upsilon) = 0,
\end{equation}
\begin{equation}
    \frac{\partial u}{\partial t} + u\frac{\partial u}{\partial R} + 
    \frac{\upsilon}{R}\frac{\partial u}{\partial \varphi} - \frac{\upsilon^{2}}{R} = 
    -\frac{\partial}{\partial R}(\mathscr{V}_{0}+\psi+h),
\end{equation}
and
\begin{equation}
    \frac{\partial \upsilon}{\partial t} + u\frac{\partial \upsilon}{\partial R} + 
    \frac{\upsilon}{R}\frac{\partial \upsilon}{\partial \varphi} + \frac{u\upsilon}{R} = 
    -\frac{1}{R}\frac{\partial}{\partial \varphi}(\psi+h),
\end{equation}
respectively, where $u(R, \varphi, t)$ and $\upsilon(R, \varphi, t)$ are the radial and
azimuthal components of velocity, $\sigma(R, \varphi, t)$ is the surface
density, $\mathscr{V}_{0}$ is the gravitational potential of SMBH, $\psi$ is the
gravitational potential of disk, $h$ is the enthalpy defined by $dh = a^2
d\sigma/\sigma$ (governed by the thermodynamic property of gas), and $a$ is the
sound speed. It should be noted that the viscosity is neglected here. 

The $m$-fold linear perturbations of the equilibrium state are considered. The
variables ($u$, $\upsilon$, $\sigma$, $\psi$, $h$) can be expressed as $F(R, \varphi,
t) = F_0(R) + F_1(R) e^{i(\omega t - m \varphi)}$, where $F$ is $u$, $\upsilon$,
$\sigma$, $\psi$, or $h$. The subscript 0 represents the variables in the
equilibrium state, and 1 represents the perturbation components. $\omega = m
\Omega_{\rm p} - i\gamma$ is the complex eigenfrequency. Its real part
represents the pattern speed $\Omega_{\rm p}$ of the rotating arms, and the
imaginary part gives the exponential growth rate $\gamma$ of the density waves.
Then, the linearized equations can be formulated as
\begin{equation}
    \frac{1}{R}\frac{d}{dR}(R\sigma_{0}u_{1})-\frac{im}{R}\sigma_{0}\upsilon_{1}+i(\omega-m\Omega)\sigma_{1}=0,
\end{equation}
\begin{equation} \label{eq:motion_1}
    i(\omega-m\Omega)u_{1}-2\Omega \upsilon_{1}=-\frac{d(\psi_{1}+h_{1})}{dR},
\end{equation}
and
\begin{equation} \label{eq:motion_2}
    \frac{\kappa^{2}}{2\Omega}u_{1}+i(\omega-m\Omega)\upsilon_{1}=im\frac{\psi_{1}+h_{1}}{R},
\end{equation}
where $\Omega(r)$ is the rotation curve and $\kappa$ is the epicyclic frequency.  

The perturbation $\psi_1$ of the gravitational potential can be given by the
integral of the surface density
\begin{equation} \label{eq:psi1}
    \psi_1(R) = -G \int_{R_{\rm in}}^{R_{\rm out}} d\zeta \int_{0}^{2\pi} 
      \frac{\zeta \sigma_1(\zeta) \cos (m \varphi) d\varphi}{\sqrt{R^2 + \zeta^2 - 2 \zeta R \cos \varphi}},
\end{equation}
where $R_{\rm in}$ and $R_{\rm out}$ are the inner and outer radius of the disk. 

Combining the above equations, we can obtain the integro-differential equation
of $\psi_1$ and $\sigma_1$
\begin{equation}\label{eq:integro-differential}
    \left[\frac{d^{2}}{dR^{2}} + {\cal A} \frac{d}{dR} + {\cal B}\right](h_1 + \psi_1) = - {\cal C} h_1,
\end{equation}
where
\begin{equation}
    {\cal A} = -\frac{d}{dR}\ln \left[\frac{\kappa^{2}(1-\nu^{2})}{\sigma_{0}R} \right], 
\end{equation}
\begin{equation}
    {\cal B} = -\frac{m^{2}}{R^{2}}-\frac{4m\Omega(R\nu^{\prime})}{\kappa R^{2}\left(1-\nu^{2}\right)} +
      \frac{2m\Omega}{R^{2}\kappa\nu}\frac{d\ln\left(\kappa^2/\sigma_{0}\Omega\right)}{d\ln R},
\end{equation}
\begin{equation}
    {\cal C} = -\frac{\kappa^{2}\left(1-\nu^{2}\right)}{a_{0}^{2}},
\end{equation}
\begin{equation}
    h_1 = a_0^2 \frac{\sigma_1}{\sigma_0},
\end{equation}
and $\nu=(\omega-m\Omega) / \kappa$ is the dimensionless frequency. Eqn
(\ref{eq:integro-differential}) is the governing integro-differential equation
of the density wave. Solving this equation numerically, if given the boundary
conditions, can provide the perturbation of the surface density $\sigma_1$.
Considering that the spiral arms in galaxies
\citep[e.g.,][]{sugai1995, aryal2004, aryal2005} or protoplanetary disks
\citep[e.g.,][]{Perez2016, huang2018} are mostly trailing (trailing waves can
transport angular momentum outward, see, e.g., \citealt{Lin1979}), we only
investigate the cases of trailing waves in BLRs in the
present paper. 

\subsection{Rotation Curve}

For the SMBH and BLR disk system, the rotation curve has three components
\citep[see][]{adams1989}
\begin{equation}
    \Omega^2(r) = \frac{G M_{\bullet}}{R^3} + \frac{1}{R}\frac{d\psi_0}{dR} + \frac{1}{R}\frac{dh_0}{dR},
\end{equation}
which come from the central SMBH, the unperturbed disk, and the pressure
respectively. The disk component can be expressed as 
\begin{equation}
    \psi_0(R) = -G \int_{R_{\rm in}}^{R_{\rm out}} d\zeta \int_{0}^{2\pi} 
      \frac{\zeta \sigma_0(\zeta) d\varphi}{\sqrt{R^2 + \zeta^2 - 2 \zeta R \cos \varphi}}. 
\end{equation}
Given the rotation curve, the epicyclic frequency can be written as 
\begin{equation}
    \kappa^2 = \frac{1}{R^3} \frac{d(R^2 \Omega)^2}{dR}.
\end{equation}
As is well known, the elliptic integral in the calculation of disk potential has
singularity \citep[e.g.,][]{adams1989, laughlin1997, hure2005}. Some methods can
handle this singularity in specific cases, e.g., the splitting method in
\cite{hure2007}. Here we follow \cite{adams1989} and use the softened gravity
method to calculate the disk potential. A softening term of $\eta^2 R^2$ is
added into the square root of the denominator at the singular points. We adopt
$\eta=0.1$ in the calculation of rotation curve, and have checked that the
deduced $\Omega(R)$ is similar to that obtained by the splitting method in
\cite{hure2007}. For Eqn (\ref{eq:psi1}), we use a smaller value of $\eta=0.01$
similar to \cite{chen2021}. We have also checked that the detailed values of the
softening parameter $\eta$ do not significantly change the spiral arms,
emission-line profiles, or velocity-delay maps in the following
sections\footnote{\cite{chen2021} did a similar check for the softened gravity
$\eta$ for protoplanetary disk in their paper, and also found that the detailed
values do not change the arms seriously.}. However, it should
be noted that the softening parameter $\eta$ may influence the growth rate of
density wave \citep[e.g.,][]{laughlin1997}, though it may not significantly
change the spiral pattern (particularly, away from the corotation or Lindblad
resonances, where $\nu=0$ or $\pm1$). We mainly focus on the spiral pattern and
the corresponding RM characteristics in the present paper. The influence of
$\eta$ to the growth rate will be discussed in future.

\subsection{Boundary Conditions}

The origin of BLRs is still under debate \citep[e.g.,][]{czerny2011, wang2017}.
Although the emissivity-averaged radii of BLRs ($R_{\rm BLR}$) have been
measured in more than 100 AGNs by RM campaigns (see, e.g., \citealt{bentz2013},
\citealt{du2015}, \citealt{du2019}, \citealt{grier2017a}), the inner/outer
radii of BLRs and their corresponding boundary conditions have large
uncertainties so far. However, the radii of dusty tori in some AGNs have been
successfully measured, which give us strong constraints to the outer radii of
their BLRs. Infrared RM campaigns found a relation between  
the radius for the innermost dusty torus and the optical luminosity, which is
written as $R_{\rm torus} \approx 0.1 L_{43.7}^{0.5} \ {\rm pc}$
\citep[e.g.,][]{minezaki2019}. $L_{43.7}$ is the V-band luminosity in units of
$10^{43.7}\ {\rm erg\ s}^{-1}$. We adopt a typical bolometric correction factor
of 10 (from bolometric to V-band luminosity). We set the outer radius of BLR at
the inner edge of dusty torus in our calculation ($R_{\rm out} = R_{\rm
torus}$). Considering $R_{\rm torus} / R_{\rm BLR} \approx 3 \sim 7$
\citep{du2015, minezaki2019}, we adopt $R_{\rm out}/R_{\rm in} = 20, 50, 100$ in
the following calculations in order to ensure that the radial range of our
calculation is wide enough, and to check the influence of different $R_{\rm
out}/R_{\rm in}$ to the spiral arms. 

We adopt the same boundary conditions as in \cite{adams1989} for simplicity, but
keep in mind that the detailed BLR boundary conditions are still unknown. At
outer boundary, the Lagrangian pressure perturbation is required to vanish,
which means the confining pressure from the external medium (probably the gas in
torus) is a constant. At the inner boundary, we assume the velocity perturbation
$u_1 = 0$, so that the radial component of the velocity
perturbation vanishes at the inner boundary.
Inner and outer boundary conditions can be verified by comparing the arm
patterns and the corresponding emission profiles, velocity-resolved lags and
velocity delay maps with the RM observations in future.

\subsection{Indirect Potential for One-armed Density Wave}

\cite{adams1989} considered the influence that the one-armed perturbation makes
the center of star be displaced from the center of mass of the protoplanetary
system for the first time. We also take this effect into account in our
calculation by the same method that incorporate an indirect potential component
in Eqn (\ref{eq:integro-differential}) as in \cite{adams1989}. The indirect 
potential can be expressed as
\begin{equation}
    \tilde{\psi}_1 = \frac{\pi \omega^2 R}{M_{\bullet} + M_{\rm disk}} \int_{R_{\rm in}}^{R_{\rm out}} 
        \zeta^2 \sigma_1(\zeta) d\zeta,
\end{equation}
where $M_{\rm disk}$ is the mass of BLR disk. 

\subsection{Numerical Method}

Exact numerical schemes for solving Eqn (\ref{eq:integro-differential}) have
been presented in, e.g., \cite{pannatoni1979} or \cite{adams1989}. In the
present paper, we adopt the matrix scheme in \cite{adams1989} for searching the
eigenvalues of $\omega$ and solving the governing integro-differential equation.
The details of the matrix scheme can be found in \cite{adams1989}. We only
briefly describe the general idea and some key points here. The integral and
differential operators in Eqn (\ref{eq:integro-differential}) can be expressed
into matrixes. By introducing the dimensionless surface density perturbation
$S(R)$ defined by $\sigma_1(R) = \sigma_0(R) S(R)$ and dividing the radial axis
to $N$ grid in logarithmic space, the integro-differential equation can be
reduced to the form of 
\begin{equation} \label{eq:matrix_N}
    \mathscr{W}_{ik}(\omega) S_k = 0,
\end{equation}
where $i, k = 1,...,N$ are the indices of the radial grid. The repeated
subscript implies summation over its range as the convention in matrix
manipulation. The first and last row of the matrix $\mathscr{W}_{ik}(\omega)$
are determined by the inner and outer boundary conditions. Eqn
(\ref{eq:matrix_N}) is a homogeneous system with $N$ equations and $N$ unknowns,
and has non-zero solutions only if the matrix $\mathscr{W}_{ik}(\omega)$ has a
vanishing determinant which can yield the eigenvalue of $\omega$.
The matrix $\mathscr{W}_{ik}(\omega)$ is a 5th-order function of $\omega$.

To find all of the eigenvalues simultaneously, Eqn (\ref{eq:matrix_N}) is 
rewritten into a $5N \times 5N$ matrix equation
\begin{equation} \label{eq:matrix_5N}
    \tilde{\mathscr{W}}_{nl}^1 S_l^* = \omega \tilde{\mathscr{W}}_{nl}^2 S_l^* ,
\end{equation}
where $n, l = 1,...,5N$ are indices, $\tilde{\mathscr{W}}_{nl}^1$ and
$\tilde{\mathscr{W}}_{nl}^2$ are two matrixes regrouped from
$\mathscr{W}_{ik}(\omega)$ in light of the coefficients of $\omega$ with
different orders, and $S_l^*$ is a rearrangement of $S_k$ (see its
detailed form in \citealt{adams1989} and Appendix \ref{app:matrix}). We can obtain the eigenvalues $\omega$ and
eigenvectors $S$ by solving this generalized eigenvalue problem. Eqn
(\ref{eq:matrix_5N}) has $5N$ eigenvalues, which is corresponding to $5N$ modes.
Most of modes have zero growth rate (imaginary part, see Section
\ref{sec:perturbation_equations} and Appendix \ref{app:eigenvalues}) and are not
physically relevant. We select the lowest order mode with significant growth
rate which will be the most global in extent and can be self-excited to become
significant. For the calculation efficiency, we use $N=500$ in the present
paper.

\section{Patterns of Spiral Arms}
\label{sec:spiral_arms}

\subsection{Fiducial Models}
\label{sec:fiducial_models}


\begin{table}
    \caption{Parameters of Models A and B}
    \label{tab:Models}
    \centering
    \begin{tabular}{cccl}
        \hline\hline
        Model      &   $p$     &   $q$      &   Note       \\
        \hline
        A & $3/4$ & $3/4$ & standard accretion disk \\
        B & $3/2$ & $1/2$ & self-gravitating  disk  \\
        \hline
    \end{tabular}
    \tablefoot{Two fiducial configurations adopted in this paper. $p$ and $q$
        are the power law indexes of surface density and sound speed,
        respectively (see Eqn \ref{eqn:surface_density} and
        \ref{eqn:sound_speed}).}
\end{table}

Before solving the governing equation, the equilibrium state of the BLR is
required. The emissivity distributions of BLRs have been preliminarily
reconstructed through BLR modeling in several objects
\citep[e.g.,][]{pancoast2012, pancoast2014, grier2017b, williams2018, li2018},
however, the real surface density distributions are still unclear because the
reprocessing coefficient distributions are not known. In Paper
\citetalias{wang2022}, we adopt the polytropic relation as the prescription of
the disk. Here, we generalize and assume that the distributions of the surface
density and sound speed are power laws, which follow
\begin{equation}
    \label{eqn:surface_density}
    \sigma_0(R) = \hat{\sigma}_0 \left(\frac{R}{R_{\rm in}}\right)^{-p},
\end{equation}
and
\begin{equation}
    \label{eqn:sound_speed}
    a_0(R) = \hat{a}_0 \left(\frac{R}{R_{\rm in}}\right)^{-q/2}.
\end{equation}
We use $q/2$ rather than $q$ as the index of $a_0$ in order to keep the same
manner as \cite{adams1989}. 

The stability of a disk can be quantified by the parameter $Q = \kappa a_0 / \pi
G \sigma_0$ \citep{toomre1964}. The disk is stable if $Q \gg 1$, and very
unstable if $Q$ is far smaller than unity. Here we consider a quasi-stable BLR
disk with the average value of $Q$ parameter, defined by
\begin{equation}
    \bar{Q} = \frac{\int_{R_{\rm in}}^{R_{\rm out}} 2 \pi R Q \sigma_0 dR}{\int_{R_{\rm in}}^{R_{\rm out}} 2 \pi R \sigma_0 dR},
\end{equation}
close to unity. We set $\bar{Q}$ as a free parameter in the following sections. 

In total, the model used here has 7 parameters: the mass of SMBH $M_{\bullet}$,
the mass ratio between disk and SMBH $M_{\rm disk} / M_{\bullet}$, the
dimensionless accretion rate \dotm, the power law indices $p$ and $q$, parameter
$\bar{Q}$, and the ratio of outer and inner radii $R_{\rm out}/R_{\rm in}$.
Among them, $M_{\bullet}$ and \dotm\ determine the outer radius, and the other 5
parameters control the pattern of spiral arms \citep[see][]{adams1989}. Changing
$M_{\rm disk} / M_{\bullet}$ is equivalent to adjusting $\sigma_0$. The value of
$\bar{Q}$ determines $a_0$ if $M_{\rm disk} / M_{\bullet}$ (equivalently
$\sigma_0$) is fixed\footnote{In practice, we adopt $Q = \Omega a_0 / \pi G
\sigma_0$ when we set the scale of $a_0$, which is similar to \cite{adams1989}.
Because the deviation of $\kappa$ from $\Omega$ in the motion equations are
treated consistently, it only makes $\bar{Q}$ slightly smaller than its true
value and doesn't influence the arm patterns and the conclusions.}. Our purpose
is not to explore the entire parameter space but to demonstrate the
observational characteristics for some typical cases of the BLR arms. Comparing
with the standard accretion disks \citep{shakura1973}, the surface density
distributions in self-gravitating accretion disks are proposed to be steeper and
$p\approx 1 \sim 3/2$ are always adopted in theoretical works
\citep[e.g.,][]{Lin1987, goodman2003}. In addition, the sound speed
distributions of self-gravitating disks are probably flatter \citep[$q=0 \sim
3/4$, see, e.g.,][]{goodman2003, sirko2003, rice2005}. We adopt ($p=3/4$,
$q=3/4$) and ($p=3/2$, $q=1/2$) as two fiducial configurations, which are
corresponding to the distributions in standard accretion disk and
self-gravitating disk, respectively. We call them Models A and B hereafter (see
Table \ref{tab:Models}). We fix $M_{\bullet}=10^8 M_{\odot}$ and $\dotm=1.0$,
and leave the other parameters ($M_{\rm disk} / M_{\bullet}$, $\bar{Q}$, and
$R_{\rm out}/R_{\rm in}$) as free parameters. $M_{\bullet}$
and $\dotm$ determine the outer radius $R_{\rm out}$. After $R_{\rm out}$ is
determined, the parameter $R_{\rm out}/R_{\rm in}$ controls the inner radius.

\begin{figure*}[!ht]
    \includegraphics[width = \textwidth]{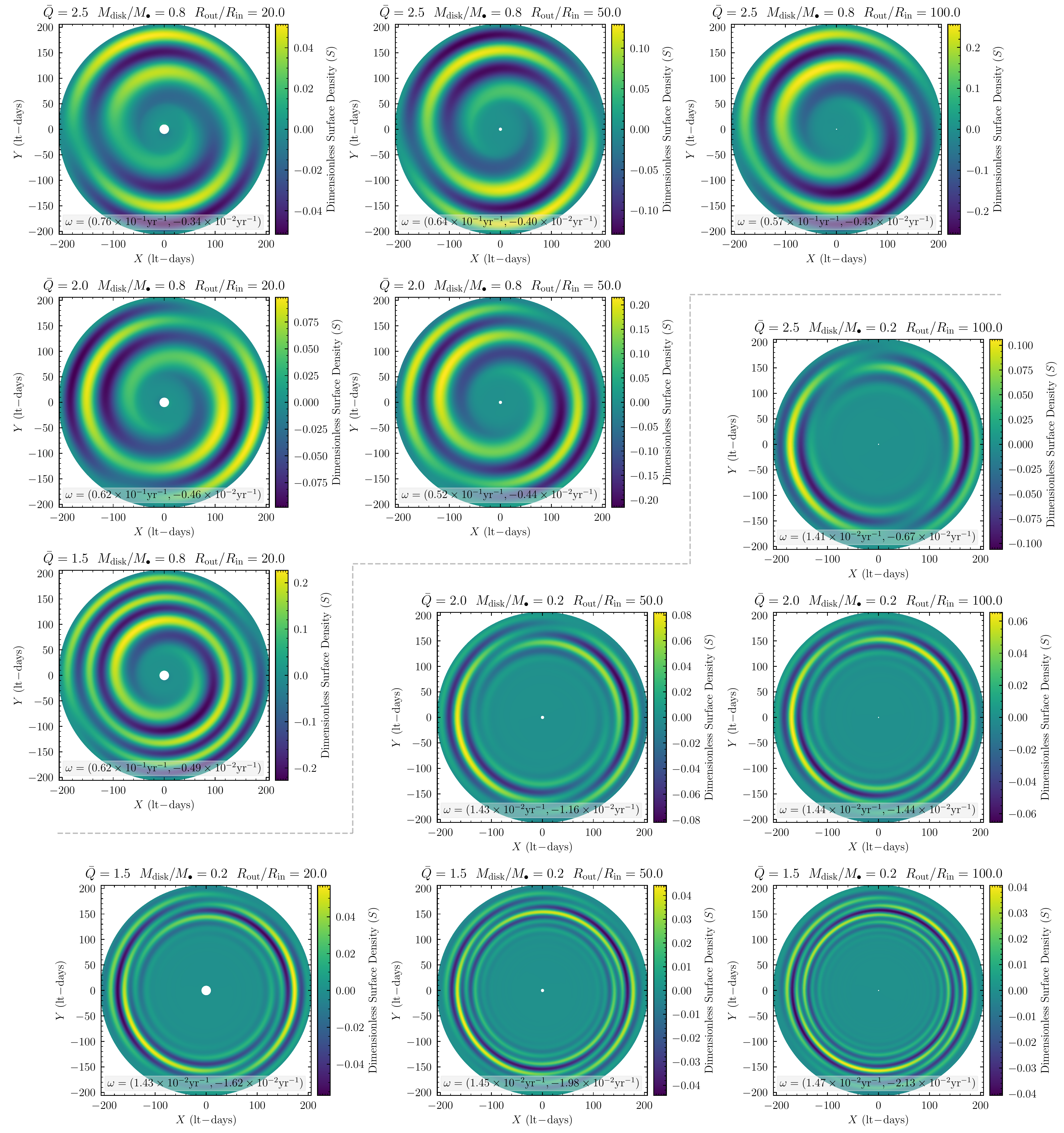}
    \caption{Dimensionless surface density of spiral arms ($m=1$) for Model A.
    The 6 panels in upper left corner are the spiral arms for more massive disks
    ($M_{\rm disk}/M_{\bullet}=0.8$), and the 6 panels in lower right corner are
    those for less massive disks ($M_{\rm disk}/M_{\bullet}=0.2$). The values of
    $\bar{Q}$, $M_{\rm disk}/M_{\bullet}$, and $R_{\rm out}/R_{\rm in}$ are
    marked on the top of each panels. In general, more massive disks have more
    loosely wound spiral arms (see more details in Section
    \ref{sec:arm_patterns}). The eigenvalues (real and imaginary parts) of
    $\omega$ are also provided in each of the panels. \label{fig:ModelAm1}}
\end{figure*}

\subsection{Spiral Arms with $m=1$}
\label{sec:arm_patterns}

Self-regulation (e.g., compression or shocks induced by the gravitational
instabilities, see \citealt{bertin1999, lodato2004, lodato2007}) has been
proposed to maintain Toomre parameter $Q$ so that it is not far smaller than
unity. In the present paper, we do not aim to investigate the detailed
self-regulation mechanisms, but simply assume that $\bar{Q}$ is a little larger
than unity \citep[see, e.g.,][]{lodato2004}. It means the disk is quasi-stable
but the instabilities can still be self-excited ($\bar{Q} = 1.5, 2.0, 2.5$). 

\begin{figure*}[!ht]
    \includegraphics[width = \textwidth]{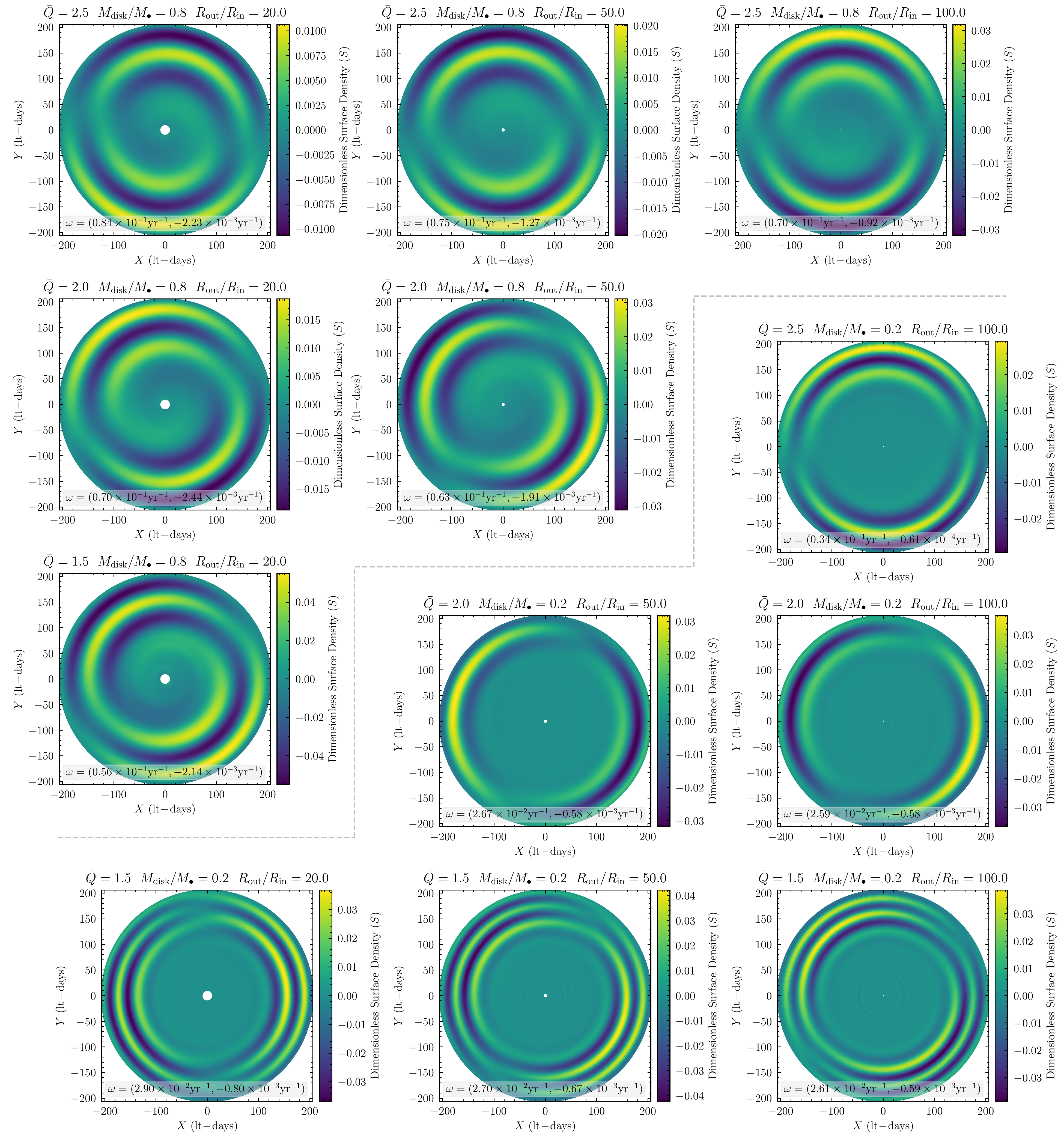}
    \caption{Dimensionless surface density of spiral arms ($m=1$) for Model B.
    Similar to Figure \ref{fig:ModelAm1}, the 6 panels in upper left corner are
    the spiral arms for more massive disks ($M_{\rm disk}/M_{\bullet}=0.8$), and
    the 6 panels in lower right corner are those for less massive disks ($M_{\rm
    disk}/M_{\bullet}=0.2$). The values of $\bar{Q}$, $M_{\rm
    disk}/M_{\bullet}$, and $R_{\rm out}/R_{\rm in}$ are marked on the top of
    each panels. The eigenvalues of $\omega$ are also provided in each of the
    panels.
    \label{fig:ModelBm1}}
\end{figure*}

It is intuitive that the one-armed density perturbation can produce the most
significantly asymmetric emission-line profiles and velocity-delay maps. We
first calculate the spiral arms of Model A with $m=1$. For each set of
parameters, there are more than one eigenvalues and solutions (modes). We adopt
the mode with the lowest order and significant growth rate because it will be
the most global and can grow in a relatively rapid rate (see the eigenvalues in
Appendix \ref{app:eigenvalues}). For $M_{\rm disk} /
M_{\bullet}$, it is still difficult to observationally determine its exact
values in AGNs, especially for the self-gravitating regions where the BLRs may
reside. But as mentioned in Section \ref{sec:Intro}, it is possible to give an
rough estimate of $M_{\rm disk} / M_{\bullet}$ from standard accretion disk
model \citep{shakura1973}, that $M_{\rm disk} / M_{\bullet}$ is in the range of
$\sim0.04-0.7$ (corresponding to $R_{\rm out}$ from $10^4R_{\rm g}$ to
$10^5R_{\rm g}$). Similarly, from the marginally self-gravitating disk model of
\cite{sirko2003}, $M_{\rm disk} / M_{\bullet}$ in quasars can be as high as a
few tenths (see Figure 2 in \citealt{sirko2003}). Here we select $M_{\rm disk} /
M_{\bullet} = 0.2$ and $0.8$ as representatives in the present paper. 
It should be noted that the disks for Model A and B are both
relatively thin with $H/R \sim 0.04$ ($M_{\rm disk}/M_{\bullet}=0.2$) and
$\sim0.15$ ($M_{\rm disk}/M_{\bullet}=0.8$), given the current disk
configurations.

The arm patterns are shown for different parameters in Figure
\ref{fig:ModelAm1}. Six cases for $M_{\rm disk} / M_{\bullet} = 0.2$ and other
six cases for $M_{\rm disk} / M_{\bullet} = 0.8$ are demonstrated (in the lower
right and upper left corners of Figure \ref{fig:ModelAm1}). Through comparing
the cases with different disk-to-SMBH mass ratios, it is obvious that more
massive disks have more loosely wound spiral arms (see more discussions in
Section \ref{sec:roles_of_Q_and_mass_ratio}). In addition, the arms in more
massive disks tend to locate in more outer radii. For the cases with the same
disk-to-SMBH mass ratio, the arms are wound more loosely if $\bar{Q}$ are larger
(see more discussions in Section \ref{sec:roles_of_Q_and_mass_ratio}). The
influence of $R_{\rm out} / R_{\rm in}$ looks very weak. 

\begin{figure*}[!ht]
    \includegraphics[width = \textwidth]{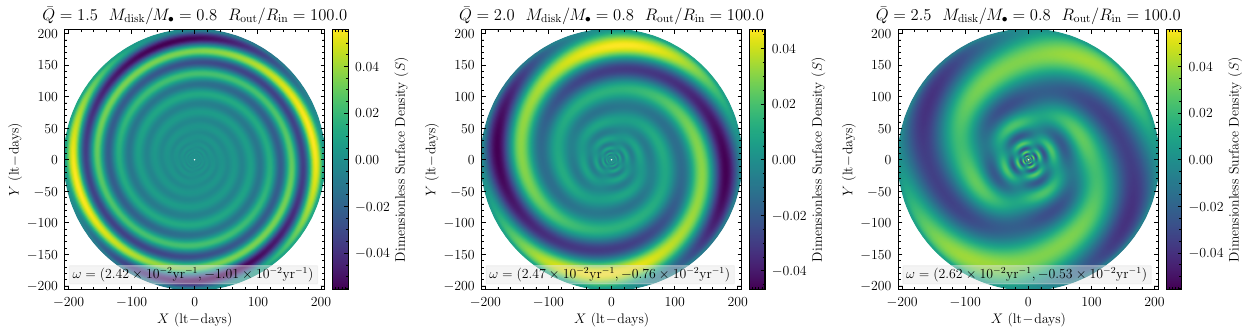}
    \includegraphics[width = \textwidth]{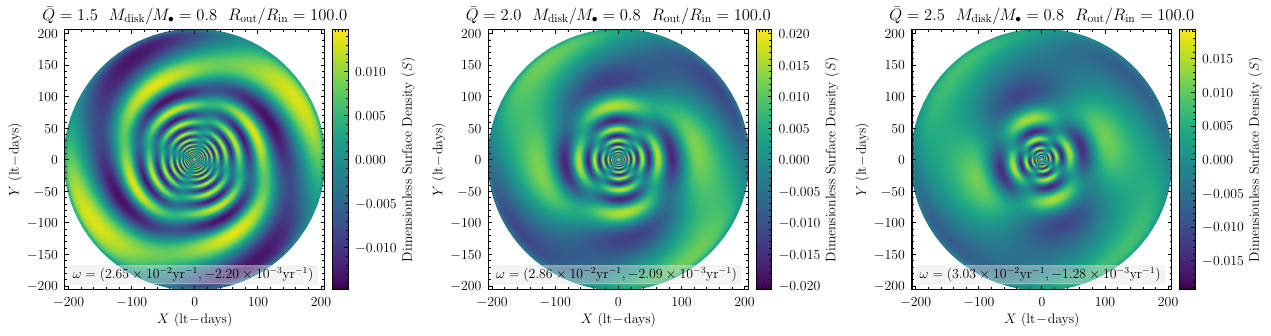}
    \caption{Dimensionless surface density of the spiral arms with $m=2$ for Models A and B. The upper 3 panels 
    are the arm patterns of Model A, and the lower 3 panels are those of Model B. We only plot the spiral arms with 
    $R_{\rm out}/R_{\rm in} = 100$ as examples. \label{fig:ModelABm2}}
\end{figure*}

We also present the spiral arms of Model B in Figure \ref{fig:ModelBm1} (the
corresponding eigenvalues are also provided in Appendix \ref{app:eigenvalues}).
In general, the spiral arms of Model B are more loosely wound than those in
Model A. Moreover, similarly, the arms in more massive disks are more loosely wound.
If $\bar{Q}$ is smaller, the spiral arms wind more tightly. The influence of
$R_{\rm out} / R_{\rm in}$ is still weak to the primary arms in the outer part
of the disk, but the inner part of the disks with larger $R_{\rm out} / R_{\rm
in}$ show some weak small arms in the less massive disks. More importantly, in
comparison with Model A, the spiral arms of Model B are more ``banana''-like
\citep[see][]{adams1989}. From the inside out, the arms in Model B do not extend
continuously but show several gaps and wiggles. In contrast, this phenomenon is
weaker in Model A. The arms in Model A extend outward more continuously. 

Our goal is to investigate the observational characteristics of the loosely
wound spiral arms. We focus on the cases with $(\bar{Q}, M_{\rm
disk}/M_{\bullet}, R_{\rm out}/R_{\rm in}) = (2.5, 0.8, 100)$ and calculate
their profiles of emission lines, the velocity-delay maps, and the
velocity-resolved lags in the following Sections \ref{sec:profiles},
\ref{sec:vdm} and \ref{sec:velocity_resolved_lags}.

\subsection{Spiral Arms with $m=2$}
\label{sec:arm_patterns_m2}

We also calculate the two-armed density waves ($m=2$). The $m=2$ spiral arms of
Models A and B with $M_{\rm disk} / M_{\bullet}=0.8$ and $R_{\rm out} / R_{\rm
in}=100$ are shown in Figure \ref{fig:ModelABm2}. Similar to $m=1$ modes, the
$m=2$ modes wind more loosely if $\bar{Q}$ is larger. Comparing
with the $m=1$ modes, the arms in the $m=2$ modes can extend inward to smaller
radii. The outer parts of the disks tend to be loosely wound, while the inner
parts wind more tightly.  In comparison with Model A, the pitch angles of the
arms in Model B are larger and the ``banana'' shape of the arms is more
significant. In the following Section \ref{sec:vdm}, we also present the
velocity-delay maps of the $m=2$ spiral arms for the cases of $(\bar{Q}, M_{\rm
disk}/M_{\bullet}, R_{\rm out}/R_{\rm in}) = (2.5, 0.8, 100)$.

\begin{table}
    \caption{Parameters of $\Xi$}
    \label{tab:reprocessing_coefficient}
    \centering
    \begin{tabular}{cccccccc}
        \hline\hline
                &    \multicolumn{3}{c}{Case I}                       &  & \multicolumn{3}{c}{Case II}                         \\  \cline{2-4} \cline{6-8}
        Model   &    $\mu_U$  & $\tilde{\sigma}_U$  & $S_{\rm max}$   &  &$\mu_U$       & $\tilde{\sigma}_U$  & $S_{\rm max}$  \\       
        \hline
            A   & 0.60        & 0.10                & 0.10            &  & 2.00         & 0.05                & 0.10           \\
            A   & 1.00        & 0.10                & 0.10            &  & 1.20         & 0.05                & 0.10           \\
            B   & 0.60        & 0.15                & 0.20            &  & 2.00         & 0.20                & 0.20           \\
            B   & 1.00        & 0.15                & 0.20            &  & 1.20         & 0.20                & 0.20           \\
        \hline
    \end{tabular}
    \tablefoot{All of the parameters are in units of $R_{\rm BLR} = 33
        \times (L_{5100} / 10^{44} \ {\rm erg \ s^{-1}})^{0.5} \ {\rm lt\!-\!days}$.
        For the typical SMBH mass $M_{\bullet}=10^8 M_{\odot}$ and accretion rate
        $\dotm=1.0$ adopted in the present paper, $R_{\rm BLR} = 40.4\ {\rm
        lt\!-\!days}$. For each case of Model A or B, we calculate the line 
        profiles for two sets of parameters in order to simulate the mean and rms
        spectra with different widths (see Section \ref{sec:profiles}). For the 
        velocity-delay maps, we adopt the same parameters for comparison.}
\end{table}

\section{Observational characteristics}
\label{sec:obs_characteristics}

\subsection{Emission-Line Profiles}
\label{sec:profiles}

\begin{figure*}[!ht]
    \includegraphics[width = \textwidth]{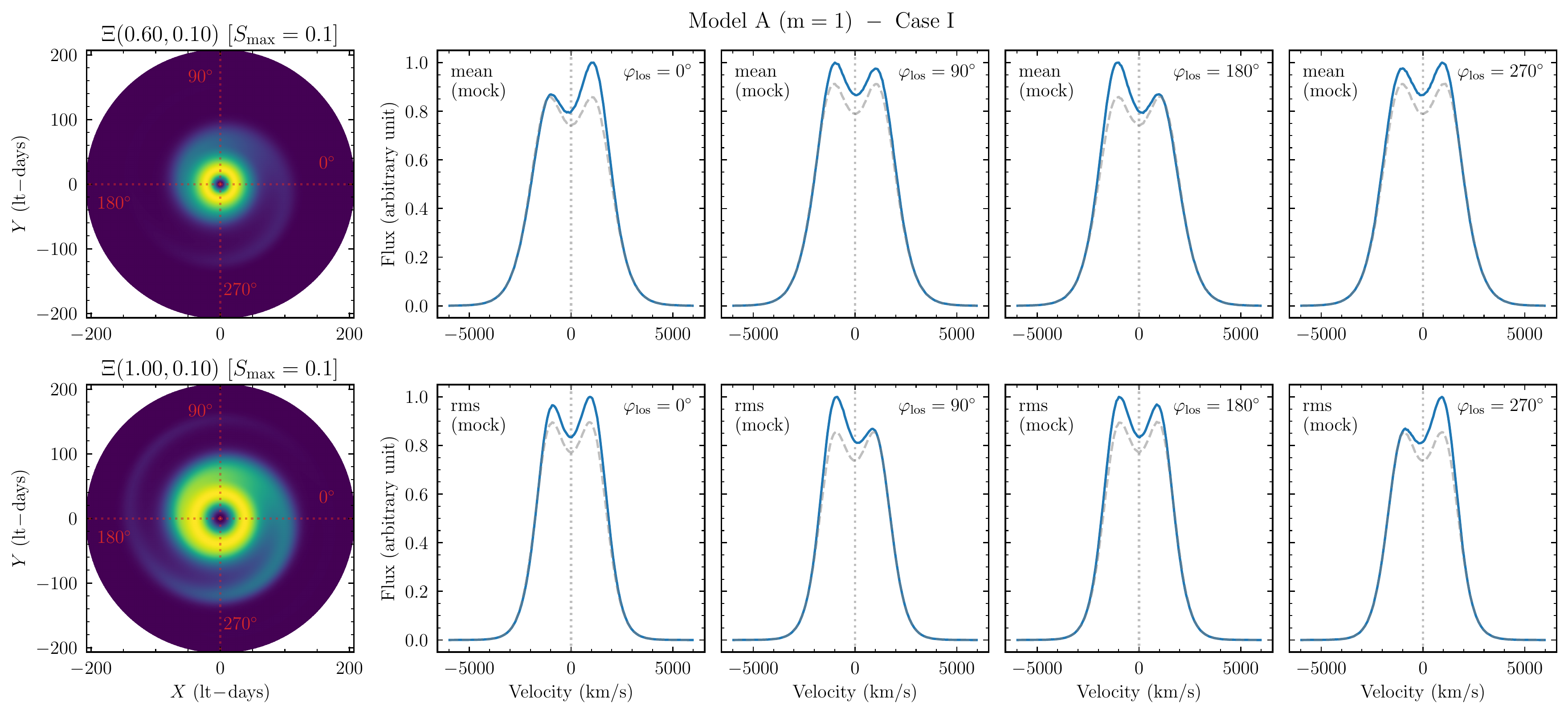}
    \includegraphics[width = \textwidth]{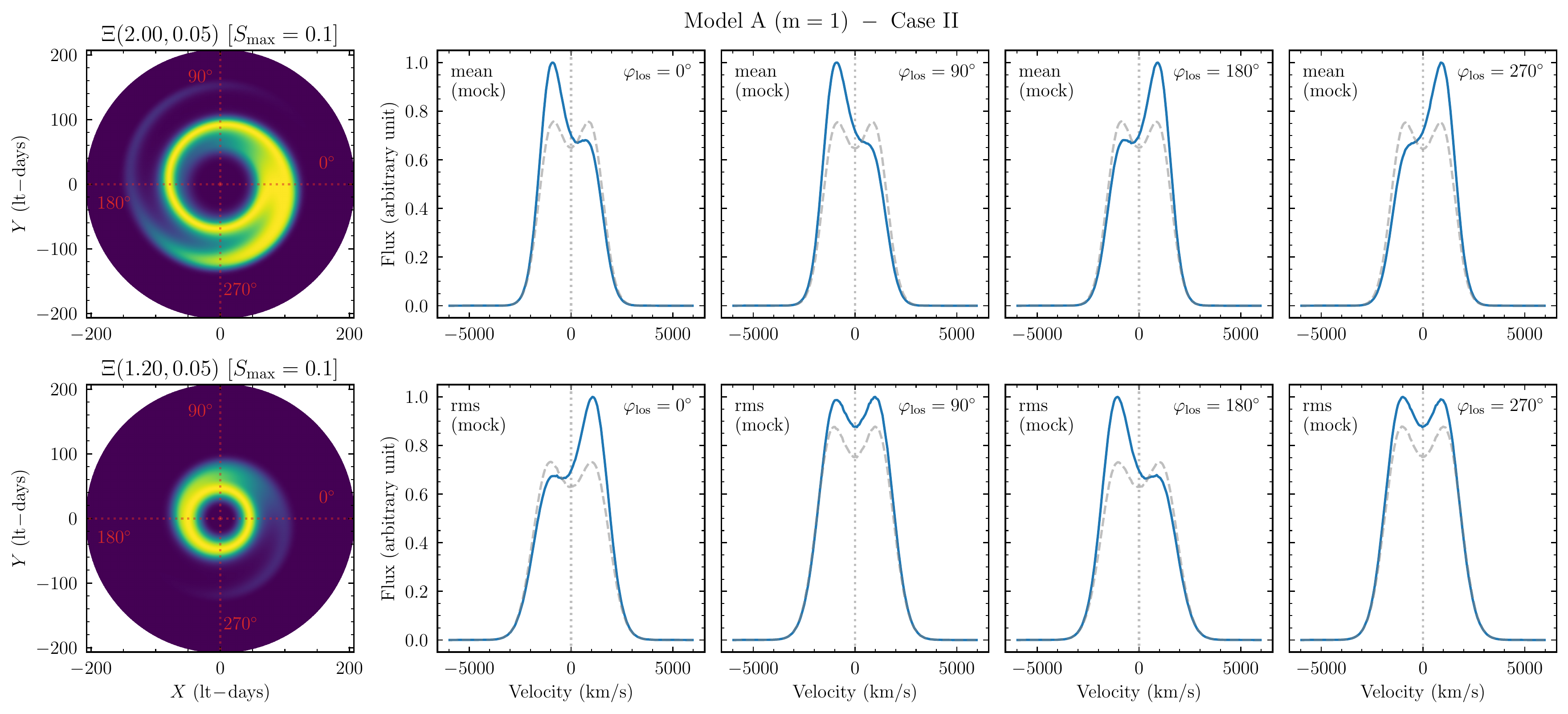}
    \caption{Emission-line profiles of Model A in Cases I and II. The left panel
    in each row is the $\Xi$ image. The values of ($\mu_U$, $\tilde{\sigma}_U$)
    and $S_{\rm max}$ are marked on the top of the $\Xi$ images. The red dotted
    lines mark the LOS azimuthal angles $\varphi_{\rm los}$. The four panels on
    the right in each row are the line profiles (blue solid lines) corresponding
    to different $\varphi_{\rm los}$. The grey dashed lines are the profiles
    without spiral arms. The line profiles (mock mean and rms) of
    the spiral arms for Case I are provided in the upper two rows, and the
    profiles of Case II are shown in the lower. \label{fig:profiles_ModelA}}
\end{figure*}

In our models, the surface density distributions are assumed to be power laws
(see Section \ref{sec:fiducial_models}). However, the emissivities of broad
emission lines do not necessarily follow the same rules. The locally optimally
emitting clouds (LOC) scenario \citep[e.g.,][]{baldwin1995, korista1997} has
been successfully applied to investigate and reproduce the observed flux ratios
of the prominent broad emission lines \citep[e.g.,][]{korista2000, leighly2004,
nagao2006, marziani2010, negrete2012, panda2018}. Its main idea is that,
although the BLR gas covers a wide range of physical conditions (e.g., density,
ionization parameter), emission lines always tend to emit from their own optimal
places \citep[e.g.,][]{baldwin1995, korista1997}. Following Paper
\citetalias{wang2022}, we simply assume that the emission-line emissivity $\Xi$
is a Gaussian function of ionization parameter $U$ of the BLR gas with the form
of
\begin{equation}
\label{eqn:Xi}
\Xi \propto \frac{1}{\sqrt{2\pi} \sigma_U} e^{-(U - U_{\rm c})^2 / 2 \sigma_U^2},
\end{equation}
where $U_{\rm c} = U(R = \mu_U R_{\rm BLR})$ is the ionization parameter corresponding to 
the most efficient reprocessing (at the radius of $R = \mu_U R_{\rm BLR}$), 
$\sigma_U = \tilde{\sigma}_U \times (U_{\rm 0,max} - U_{\rm 0,min})$ represents the range of efficient reprocessing, 
$U_{\rm 0,max}$ and $U_{\rm 0,min}$ are the maximum and minimum ionization parameters in the unperturbed disk, 
$\mu_U$ and $\tilde{\sigma}_U$ are two dimensionless parameters. The ionization parameter of the BLR gas
is defined as
\begin{equation}
    U = \frac{Q_{\rm H}}{4 \pi R^2 c n_{\rm H}},
\end{equation}
where $Q_{\rm H}$ is the number of hydrogen-ionizing photons, $n_{\rm H} = \rho
/ m_{\rm H}$ is the number density, $\rho = (\sigma_0 + \sigma_1) / 2H =
(\sigma_0 + \sigma_1) \Omega / 2 a_0$ is the hydrogen density, and $m_{\rm H}$
is the mass of hydrogen. The line profile can be expressed as 
\begin{equation}
    F_{\ell}(\lambda) = \! \! \int_{R_{\rm in}}^{R_{\rm out}} \! \! \! \! R dR 
    \int_{0}^{2 \pi} \! \! \! \Xi g(R, {\bm \upsilon}) \,
    \delta \! \left[\lambda - \lambda_0 \! \left( \! 1 + \frac{{\bm \upsilon} 
    \cdot {\bm n}_{\rm obs}}{c} \! \right) \right] d\varphi,
\end{equation}
where $\lambda_0$ is the central wavelength of the emission line, ${\bm \upsilon}(R,
\varphi)$ is the velocity of the BLR gas, $g(R, {\bm \upsilon})$ is the velocity
distribution at $R$, and ${\bm n}_{\rm obs}$ is the unit vector pointing from
the observer to the source (LOS).

\begin{figure*}[!ht]
    \includegraphics[width = \textwidth]{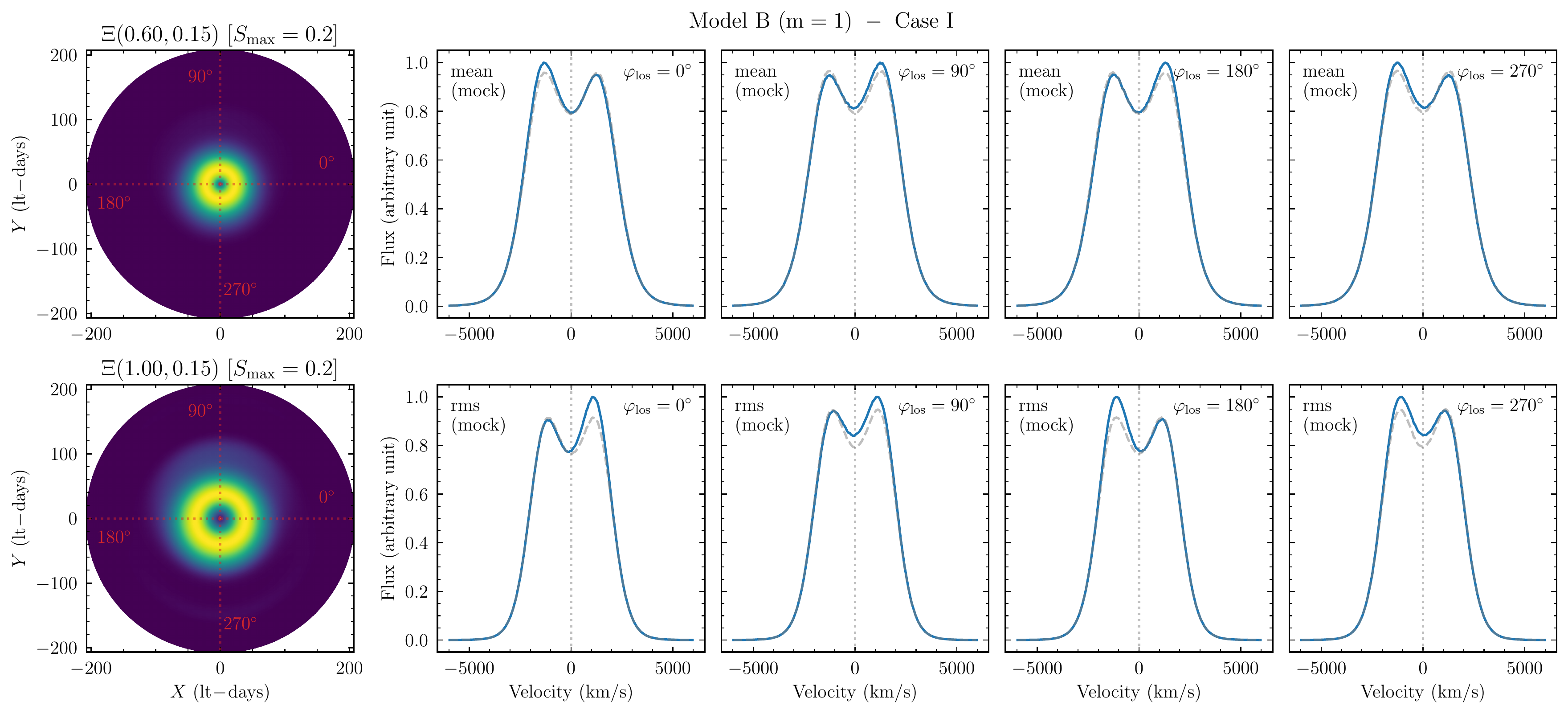}
    \includegraphics[width = \textwidth]{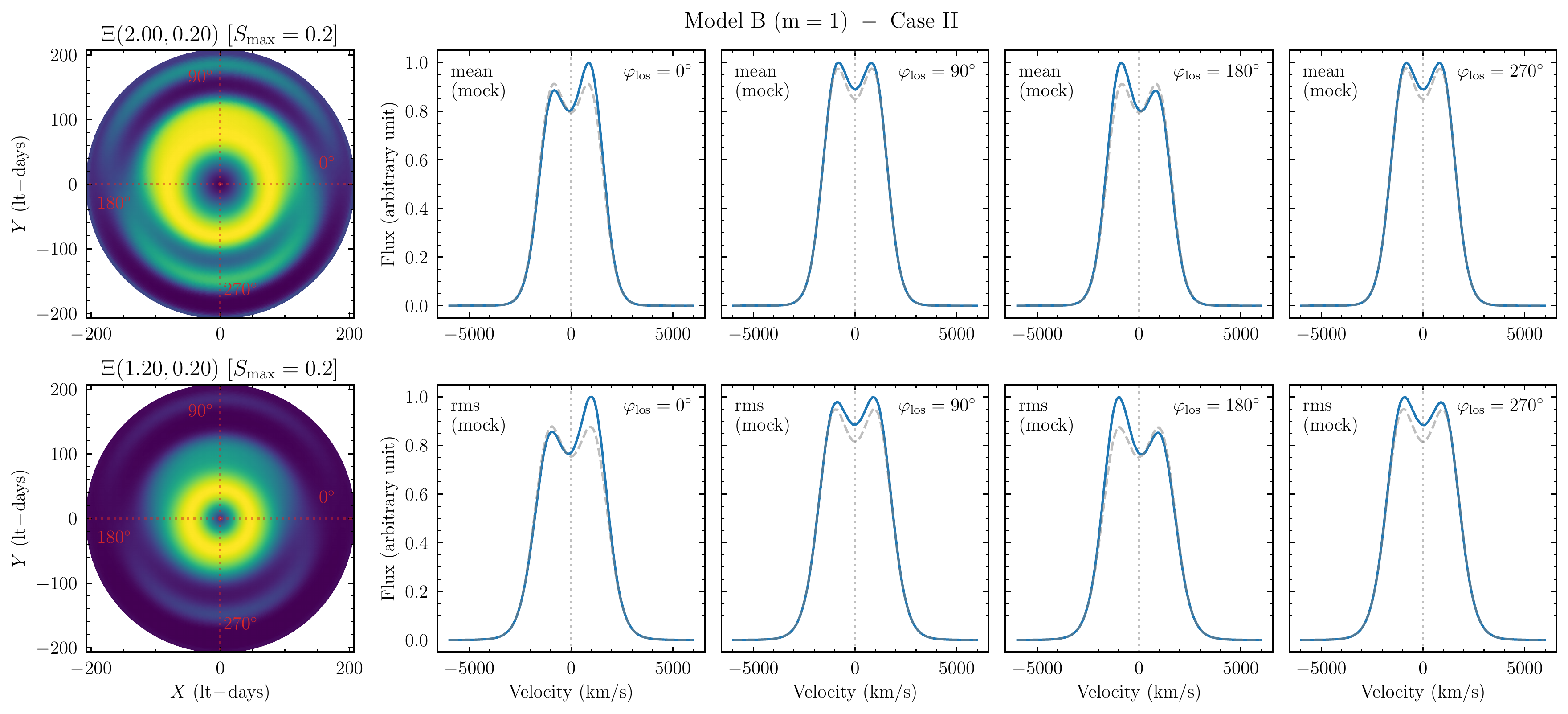}
    \caption{Emission-line profiles of Model B in Cases I and II. The meaning of
    panels and different lines (solid and dashed) are the same as in Figure
    \ref{fig:profiles_ModelA}. \label{fig:profiles_ModelB}}
\end{figure*}

Many RM campaigns have demonstrated that their rms spectra have line widths
different (narrower or broader) from the mean spectra
\citep[e.g.,][]{peterson1998, bentz2009, denney2009, barth2013, fausnaugh2017,
grier2012, du2018b, derosa2018, brotherton2020}, which means the responsivity
(varying part) of BLR is different from its mean emissivity. For simplicity, to
simulate this phenomenon, we simply assume the responsivity has the same form as
Eqn (\ref{eqn:Xi}) but with a different set of ($\mu_U$, $\tilde{\sigma}_U$)
rather than taking into account the real photoionization processes in our
calculation (hereafter we use $\Xi$ to denote both of emissivity and
responsivity). Here we investigate two combinations of ($\mu_U$,
$\tilde{\sigma}_U$) corresponding to the typical cases that rms spectra are
narrower or broader (called Cases I and II). The values of $\mu_U$,
$\tilde{\sigma}_U$, and the maximum dimensionless surface density $S_{\rm max}$
are listed in Table \ref{tab:reprocessing_coefficient}. We select these
parameters because on one hand they can demonstrate the line profiles (or
velocity-delay maps in the following Section \ref{sec:vdm}) at different radii,
and on the other hand it is easy for us to simulate the mean and rms spectra
with different line widths. We set the maximum value of dimensionless surface
density $S_{\rm max}$ to 0.1 and 0.2 for Models A and B, respectively (also in
the following Sections \ref{sec:vdm} and \ref{sec:velocity_resolved_lags}). It
should be noted that the actual situations may probably be larger or smaller
than these values. More detailed calculations including photoionization models
will be carried out in a separate paper in future.

\subsubsection{Line Profiles with $m=1$}
\label{sec:profiles_m1}

We present the emission-line profiles of single-epoch/mean spectra and rms
spectra for the spiral arms of Models A with $m=1$, for different azimuthal
angles ($\varphi_{\rm los}$) of LOS, in Figure \ref{fig:profiles_ModelA}. The
disks are rotating counter-clockwise. The LOS inclination angle only changes the
widths of emission lines, we fix the inclination angle to $\theta_{\rm
los}=30^{\circ}$ in our calculation ($\theta_{\rm los}=0^{\circ}$ refers to
looking at the disks from the face-on direction).  The contribution of the sound
speed $a_0$ is also taken into account by adding a macro-turbulence speed in the
direction perpendicular to the disk. For each of Cases I and II, the mock mean
and rms spectra are provided as two rows in Figure \ref{fig:profiles_ModelA}. As
expected, the mean spectra are broader than the rms spectra in Case I, and are
relatively narrower in Case II. It is obvious that the line profiles are {\rm
generally} double-peaked because the most efficient emitting region resemble a
ring-like shape (determined by Eqn (\ref{eqn:Xi})). The stonger
emissivities/responsivities of the spiral arms results in an obvious asymmetry
in the line profiles (see Figure \ref{fig:profiles_ModelA}). Along with the
azimuthal angle $\varphi_{\rm los}$ increases from $0^{\circ}$, the asymmetry of
the profiles change between symmetric, blueward, and redward periodically. For
some cases, the weaker peaks almost disappear (e.g., $\varphi_{\rm
los}=90^{\circ}$ in the first row of Case II). In Case II, the asymmetries
caused by the spiral arms are more significant because the $\mu_U$ parameters
are relatively larger and the $\tilde{\sigma}_U$ are smaller. More importantly,
the asymmetries of the mock mean and rms spectra can be totally different
(blueward or redward) even if the LOS are exactly the same (see, e.g.,
$\varphi_{\rm los}=180^{\circ}$ in Case II). It implies that the spiral arms can
naturally produce differently-asymmetric mean and rms spectra without any
further special assumptions. 

In Model B, its emissivity/responsivity tends to be distributed in more outer
radius (because $U \propto R^{3/4}$ approximately). The
emissivities/responsivities of the spiral arms and the corresponding
emission-line profiles for Model B in Cases I and II are shown in Figure
\ref{fig:profiles_ModelB}. The ``banana''-like distributions of the spiral arms
in Model B (see Section \ref{sec:arm_patterns} and Figure \ref{fig:ModelBm1})
still make the emission-line profiles significantly asymmetric. Compared with
Model A, Model B has relatively less asymmetric line profiles.

Some of the mock line profiles in Figures \ref{fig:profiles_ModelA} and
\ref{fig:profiles_ModelB} are very similar to the observations. We will provide
a simple comparison between the models and observations in the following Section
\ref{sec:compare_profiles}.

\subsubsection{Line Profiles with $m=2$}
\label{sec:profiles_m2}
    
For the spiral arms with $m=2$, the profiles of their corresponding emission
lines are symmetric and double-peaked. The perturbation $\sigma_1$ is identical
if $\varphi$ increases every $180^{\circ}$ ($m$-fold axis-symmetric), so the
emissivities on the left and right sides of the LOS (blueshifted and redshifted)
are exactly the same. Therefore, the line profiles of the arms with $m=2$ have
no asymmetry. The readers can refer to the dashed lines in Figures
\ref{fig:profiles_ModelA} and \ref{fig:profiles_ModelB}.

\begin{figure*}[!ht]
    \includegraphics[width = \textwidth]{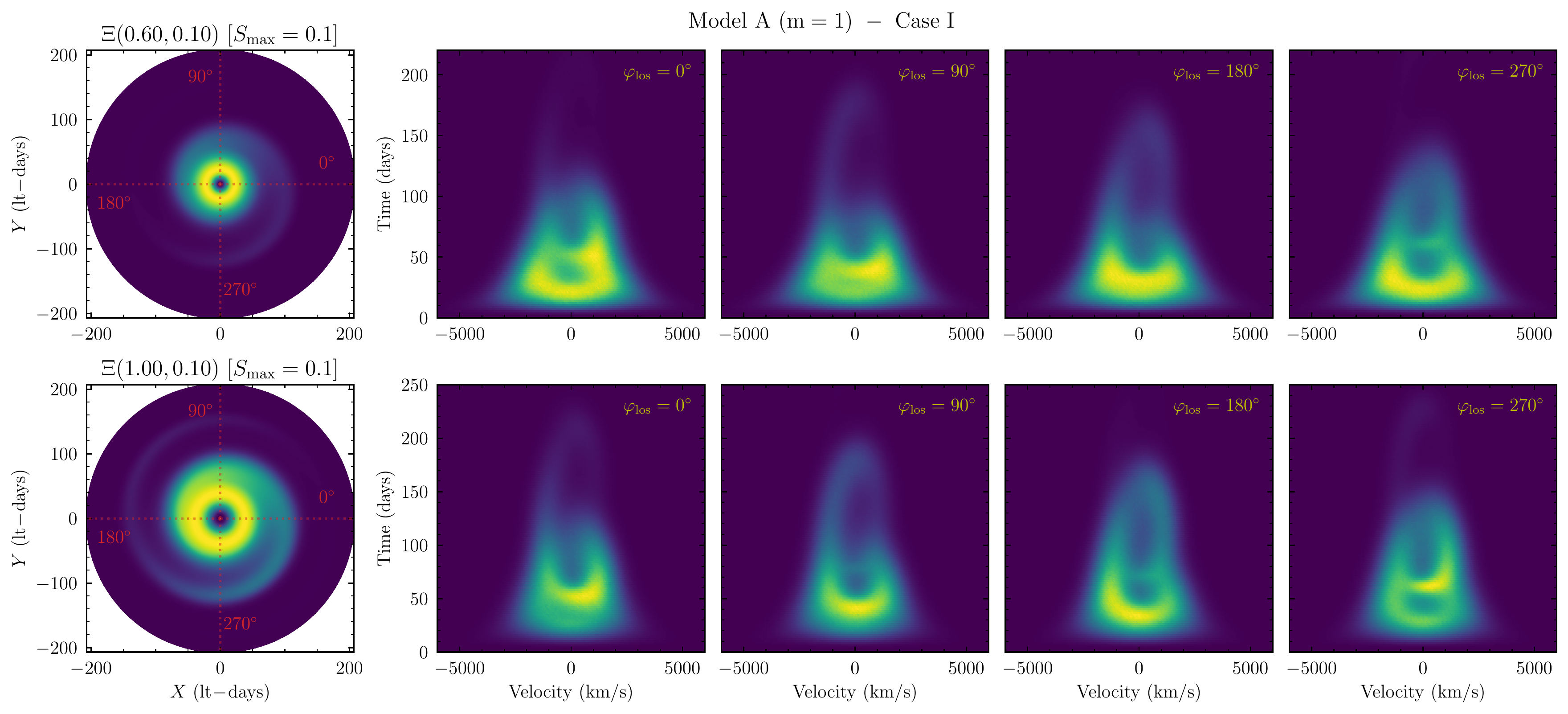}
    \includegraphics[width = \textwidth]{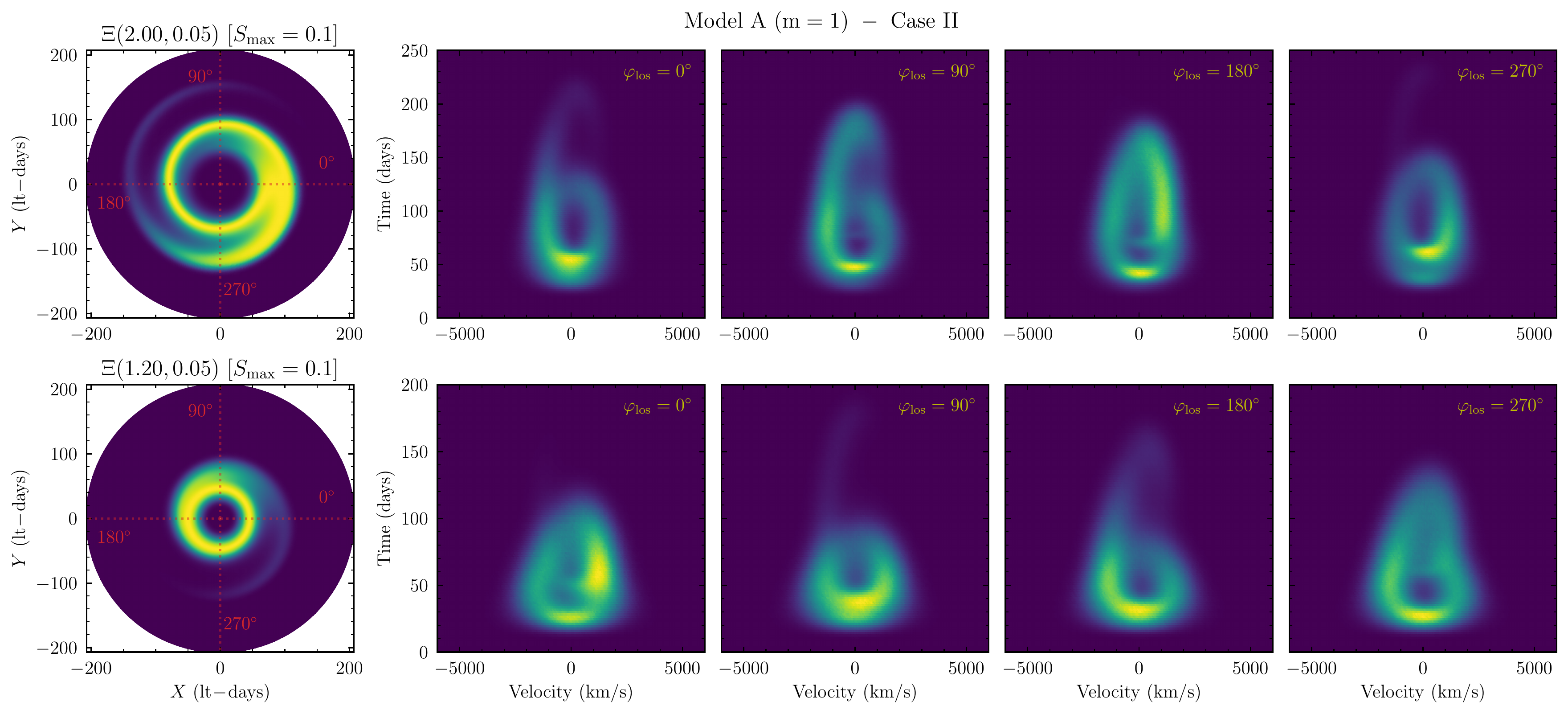}
    \caption{Velocity-delay maps of Model A ($m=1$) in Cases I and II. The left panel in
    each row is the $\Xi$ image. The red dotted lines
    mark the LOS azimuthal angles $\varphi_{\rm los}$. The four panels on
    the right in each row are the velocity-delay maps corresponding
    to different $\varphi_{\rm los}$ \label{fig:vdms_ModelA}}
\end{figure*}

\subsection{Velocity-delay Maps}
\label{sec:vdm}

RM can be approximated as a linear model that 
\begin{equation}
    \Delta L_{\rm \ell} (\upsilon, t) = \int_{-\infty}^{+\infty} \Psi(\upsilon, \tau) \Delta L_{\rm c}(t - \tau) d\tau,
\end{equation}
where $\Psi(\upsilon, \tau)$ is the so-called ``velocity-delay map'' (or
transfer function), $\Delta L_{\rm c}(t)$ is the continuum light curve, and
$\Delta L_{\rm \ell} (\upsilon, t)$ is the variation of emission-line profile at
different epochs \citep[e.g.,][]{blandford1982}. The velocity-delay map
describes how the line profile responses to the varying continuum flux, and is
determined by the geometry, kinematics and emissivity of the gas in BLR.  
The velocity-delay map of a simple Keplerian disk is symmetric, and has been
calculated numerically and demonstrated in many works (\citealt{welsh1991,
perez1992, horne2004, grier2013}, or see Appendix D in Paper
\citetalias{wang2022}). 

The velocity-delay map can be calculated from
\begin{equation}
    \begin{split}
    \Psi(\upsilon, t) = \int_{R_{\rm in}}^{R_{\rm out}}  R dR 
    \int_{0}^{2 \pi} \Xi g(R, {\bm \upsilon}) 
    \delta (\upsilon - \bm{\upsilon} \cdot \bm{n}_{\rm obs}) \\
    \times  \delta \left[ t - \frac{R + \bm{R} \cdot \bm{n}_{\rm obs}}{c} \right] d\varphi.
    \end{split}
\end{equation}
In the calculation of emission-line profiles, we adopted two sets of
parameters ($\mu_U$, $\tilde{\sigma}_U$) for each case in Models A and B in
order to simulate the mean and rms spectra ($\Xi$ represents emissivity
and responsivity, respectively). Strictly speaking, in the calculation of
velocity-delay maps, we ought to employ the ``responsivity'' implication of $\Xi$,
however, we do not distinguish responsivity and emissivity here because
we simply assumed that they have the same form mathematically (Gaussian
distributions, see Section \ref{sec:profiles}). The only difference between
them is that their ($\mu_U$, $\tilde{\sigma}_U$) are
not the same, which means the BLR gas with most efficient responses/emissivities
are located at different radii. We still calculate the velocity-delay maps using
the $\mu_U$ and $\tilde{\sigma}_U$ in Table \ref{tab:reprocessing_coefficient}, and
use the nomenclature $\Xi$ in the following discussions. The LOS
inclination angle is fixed to $\theta_{\rm los}=30^{\circ}$. A smaller or larger
angle will cause the velocity-delay maps narrow or broader in their velocity
axes. 

\subsubsection{Velocity-delay Maps with $m=1$}
\label{sec:vdms_m1}

\begin{figure*}[!ht]
    \includegraphics[width = \textwidth]{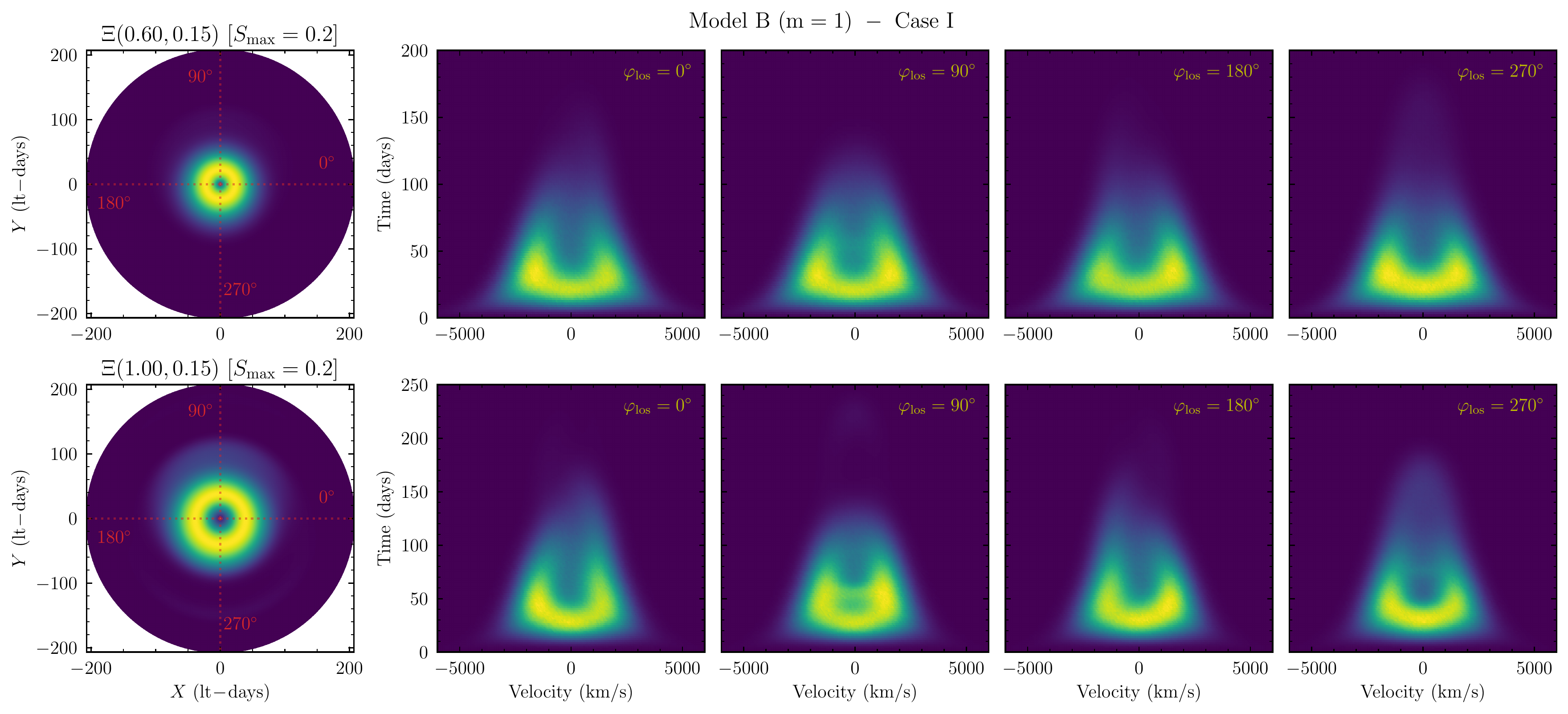}
    \includegraphics[width = \textwidth]{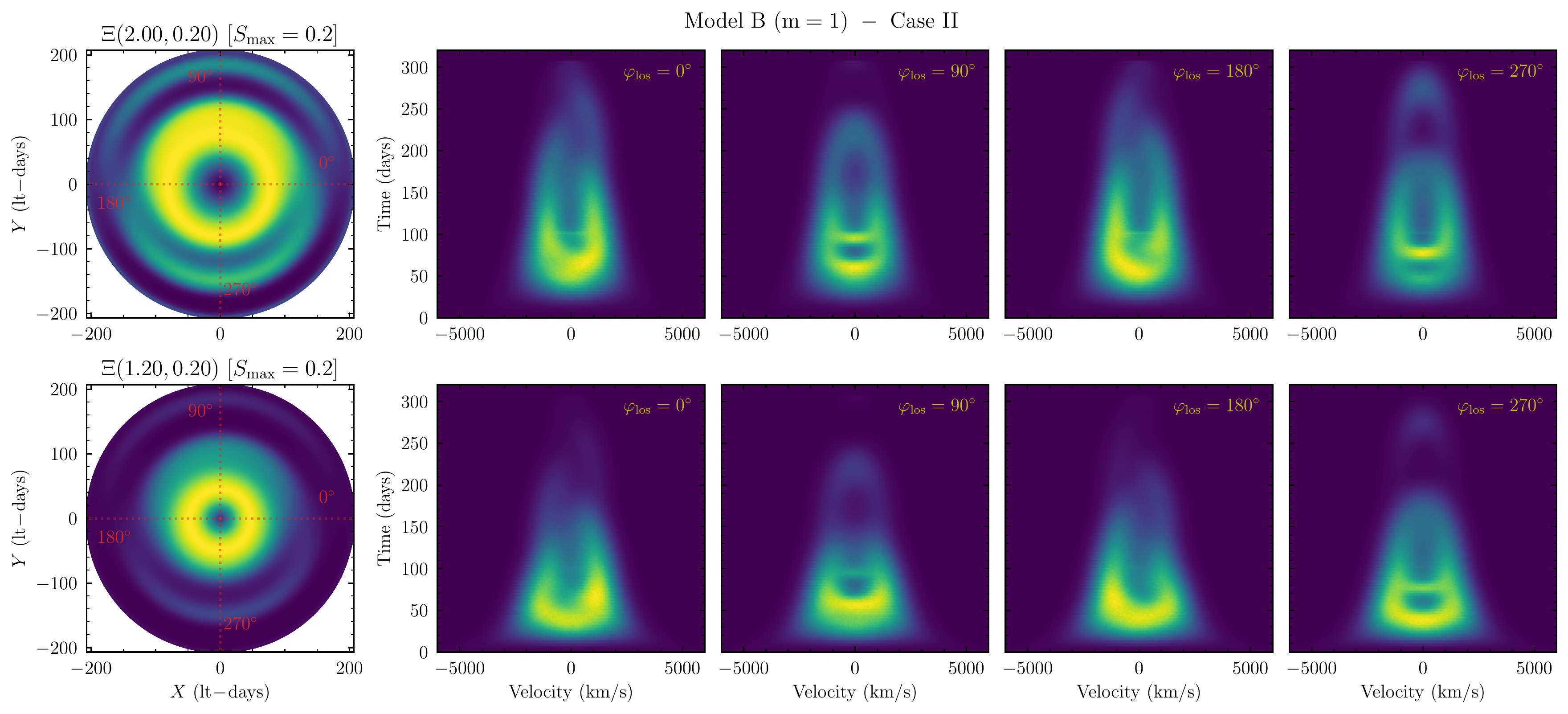}
    \caption{Velocity-delay maps of Model B ($m=1$) in Cases I and II. The meaning of
    panels are the same as in Figure \ref{fig:vdms_ModelA}.
    \label{fig:vdms_ModelB}}
\end{figure*}

Similar to the line profiles, we calculate the velocity-delay maps of Models A
and B for different LOS azimuthal angles. The results for both of Cases I and II
are provided (see Figures \ref{fig:vdms_ModelA} and \ref{fig:vdms_ModelB}). The
sound speed has also been taken into account, so the corresponding
velocity-delay maps look moderately smooth. The general morphologies of the
velocity-delay maps are similar to the traditional ``bell''-like envelope with a bright
``elliptical ring'' of a simple Keplerian disk \citep[e.g.,][]{welsh1991,
perez1992, horne2004, grier2013}. However, they are significantly asymmetric and show
remarkable sub-features of bright arcs/strips (indicating strong 
responses from the arms). The asymmetries of the responses in the velocity-delay maps are
consistent with the asymmetries of the line profiles in Figures
\ref{fig:profiles_ModelA} and \ref{fig:profiles_ModelB}.

In Model A, the contributions from the strong responsivities of the spiral arms
look significant (see Figure \ref{fig:vdms_ModelA}). Along with the azimuthal
angle increases from $0^{\circ}$ to $270^{\circ}$, the asymmetry and the
locations of the arcs/strips in the maps caused by the strong arm responsivities
change correspondingly. 

In Case II, the spiral arm patterns are more significant in the $\Xi$
distributions if the strong-response regions are mainly located in larger radii.
The bright arcs/strips (the strongest responses) in the velocity-delay maps are
corresponding to the crests of the density waves. For $\mu_U=1.20$,
$\tilde{\sigma}_U=0.05$, and $\varphi_{\rm los}=270^{\circ}$, the emission-line
profile in Figure \ref{fig:profiles_ModelA} is almost symmetric and
indistinguishable from that of a simple Keplerian disk. However, the
velocity-delay map can break this degeneracy. The sub-features in the
corresponding velocity-delay map are obvious and asymmetric distributed. In the
velicity-delay map of $\mu_U=0.60$, $\tilde{\sigma}_U=0.10$ (or $\mu_U=1.00$,
$\tilde{\sigma}_U=0.10$), and $\varphi_{\rm los}=90^{\circ}$, there is an arc
that starts from blue velocities and extends toward long time lag. But the arc
doesn't circle back to the red velocities.  It is very interesting that this
sub-feature is almost the same as the observation of NGC 5548 (incomplete
ellipse, see Figure 3 in \citealt{xiao2018b} and Figure 5 in
\citealt{horne2021}).

For Model B, the morphologies of the responsivity ($\Xi$) distributions are more
``banana''-like (bright on one side, and dark on the opposite side). The
asymmetries and sub-structures in the velocity-delay maps are a little weaker
(but still significant) than Model A. The semicircle arcs in $\Xi$ (see Figure
\ref{fig:vdms_ModelB}) result in strips and arcs overlapped with the original
``bell''-like signatures in the velocity-delay maps. These sub-features (bright
arcs and strips) rotate close-wise along with the LOS azimuthal angle increases.

\begin{figure*}[!ht]
    \includegraphics[width = \textwidth]{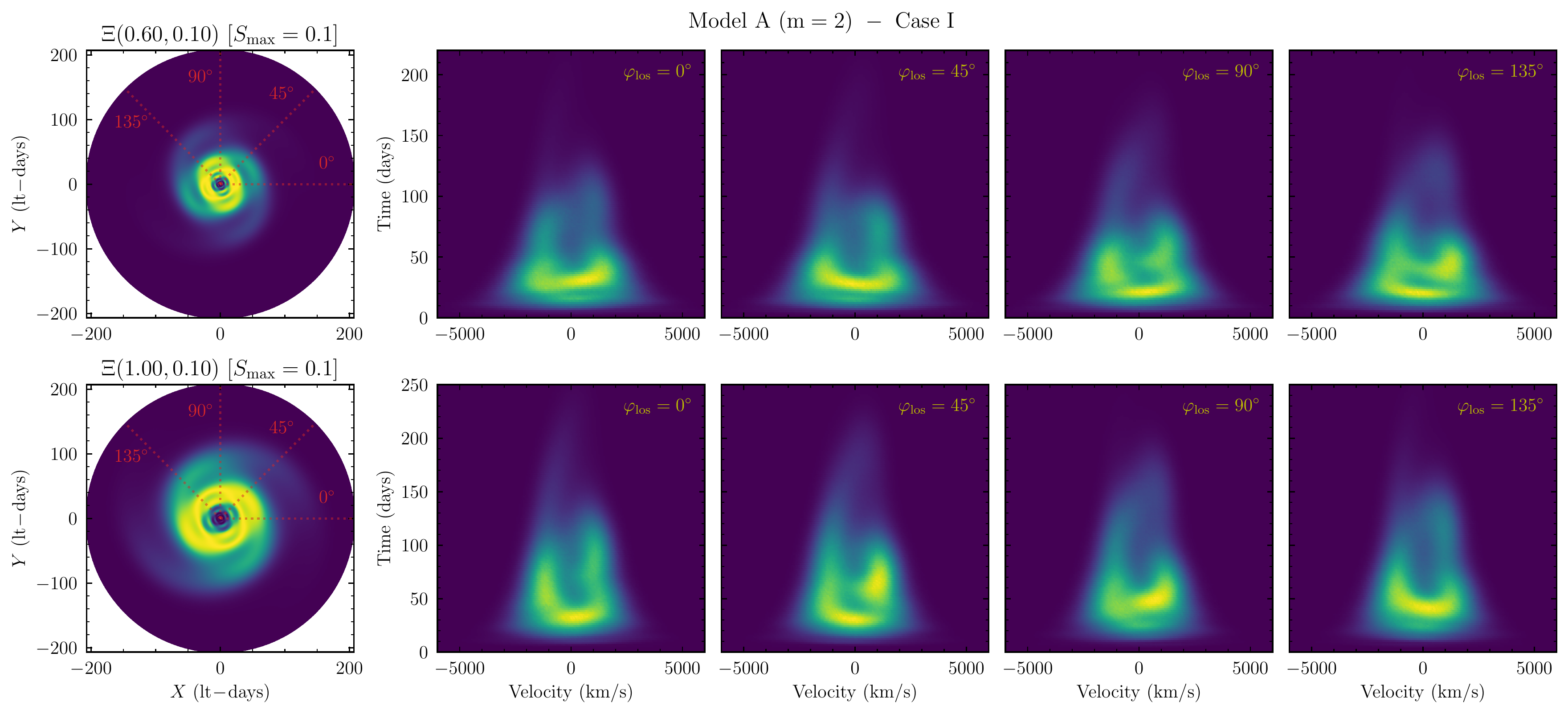}
    \includegraphics[width = \textwidth]{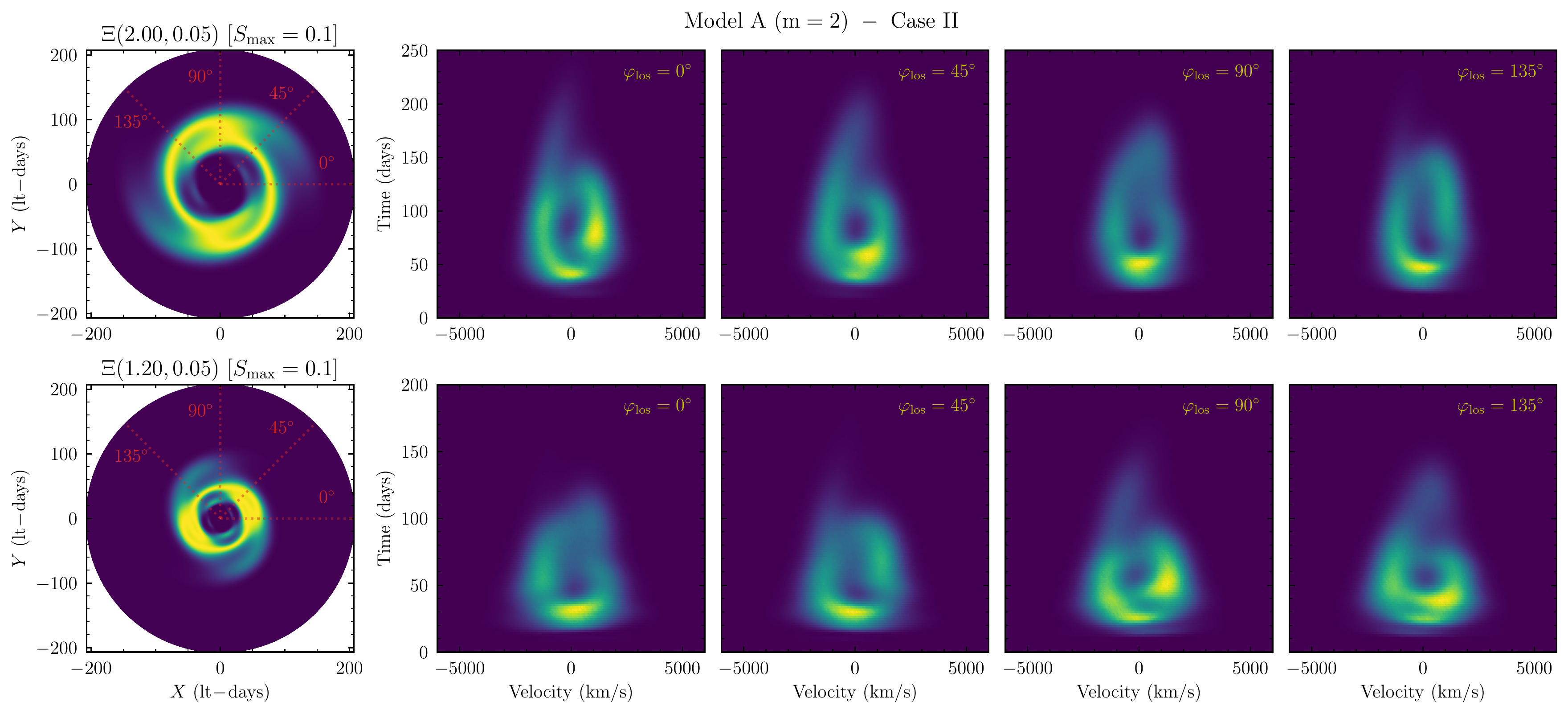}
    \caption{Velocity-delay maps of Model A ($m=2$) in Cases I and II. The meaning of
    panels are the same as in Figure \ref{fig:vdms_ModelA}.
    \label{fig:vdms_ModelAm2}}
\end{figure*}

\subsubsection{Velocity-delay Maps with $m=2$}
\label{sec:vdms_m2}

\begin{figure*}[!ht]
    \includegraphics[width = \textwidth]{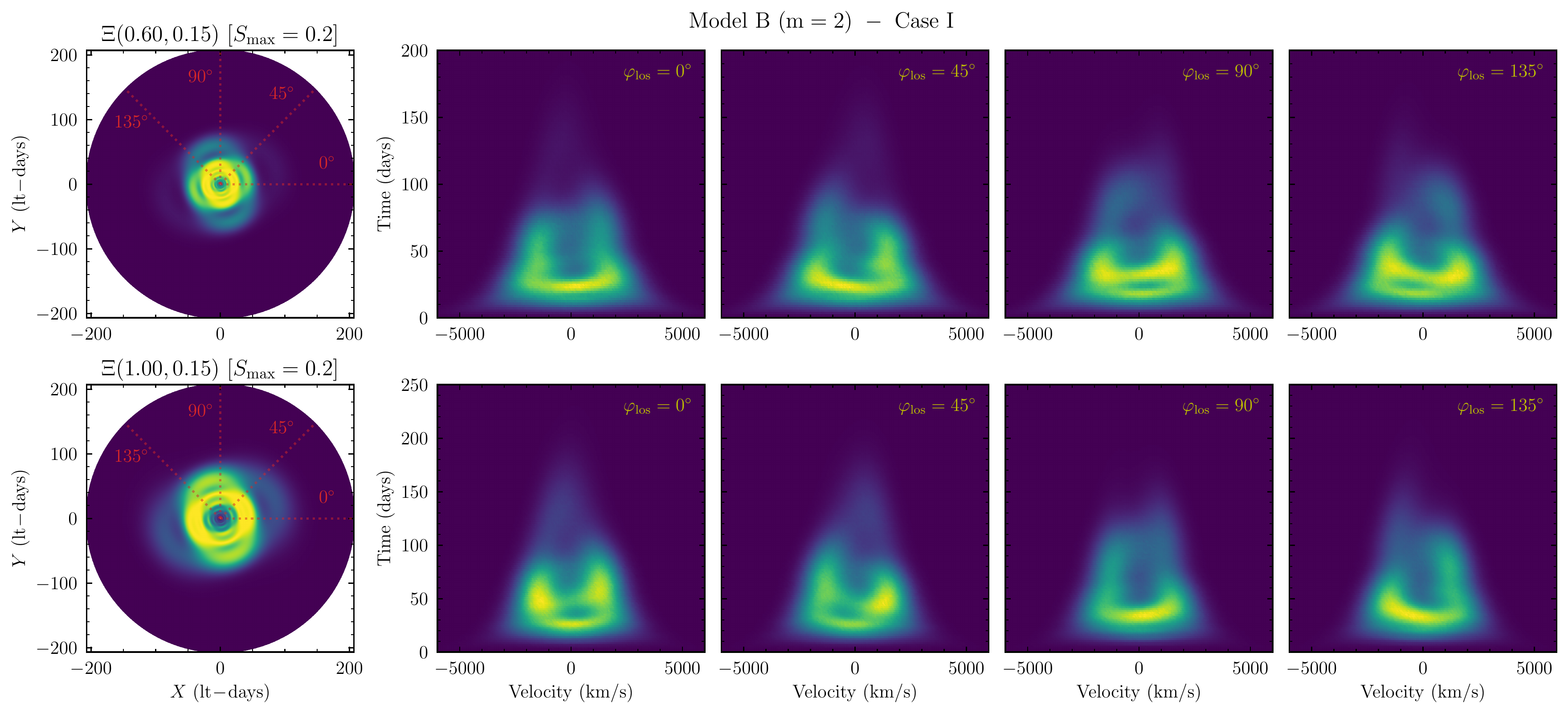}
    \includegraphics[width = \textwidth]{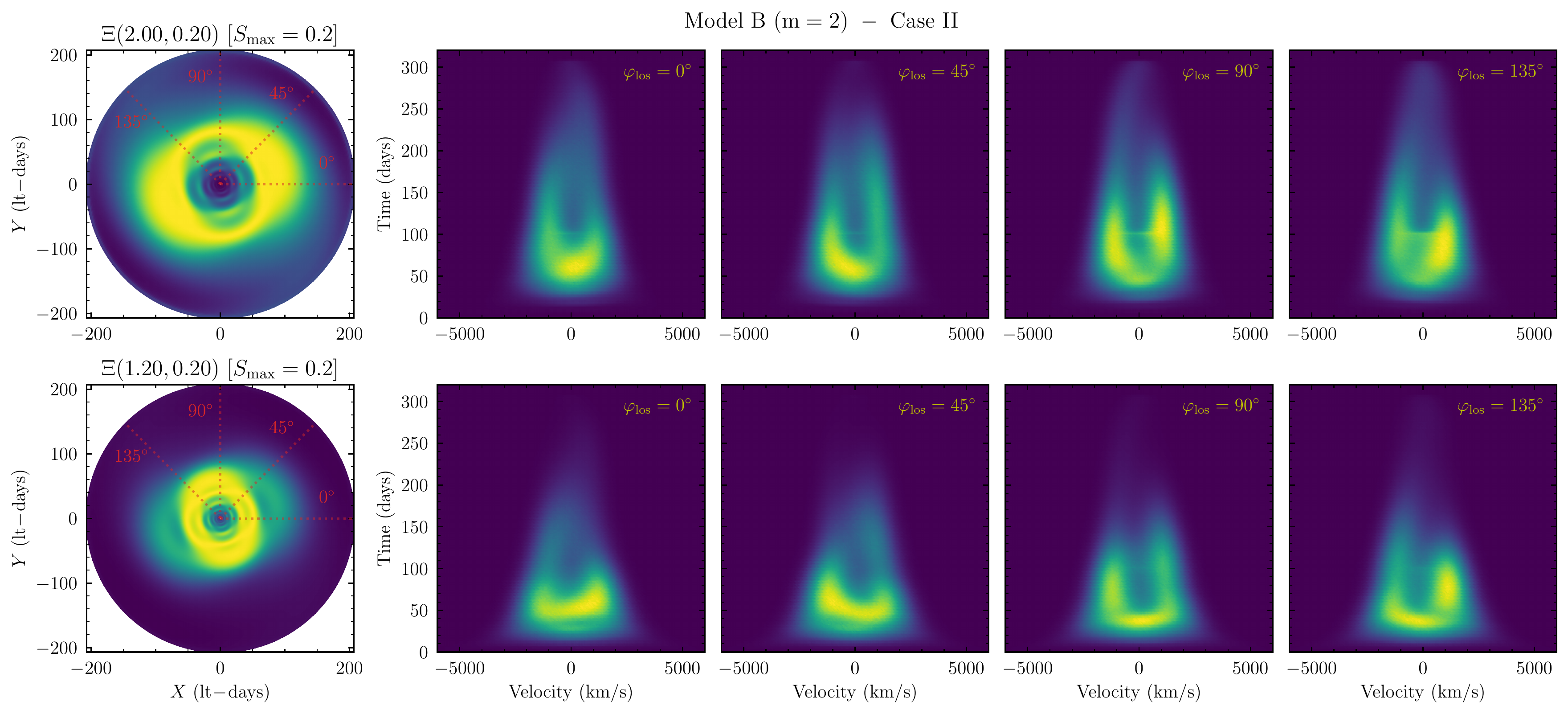}
    \caption{Velocity-delay maps of Model B ($m=2$) in Cases I and II. The meaning of
    panels are the same as in Figure \ref{fig:vdms_ModelBm2}.
    \label{fig:vdms_ModelBm2}}
\end{figure*}

As mentioned in Section \ref{sec:profiles_m2}, the line profiles of the spiral
arms with $m=2$ have no asymmetries and are not different from the profiles of a
simple Keplerian disk. However, the velocity-delay maps can break this
degeneracy. The maps of the $m=2$ arms have significant sub-features and may be
distinguishable in observations. We calculate the corresponding velocity-delay
maps for Models A and B with $m=2$ (see Figures \ref{fig:vdms_ModelAm2} and
\ref{fig:vdms_ModelBm2}). Similar to Section \ref{sec:vdms_m1}, we adopt
$(\bar{Q}, M_{\rm disk}/M_{\bullet}, R_{\rm out}/R_{\rm in}) = (2.5, 0.8, 100)$.
The spiral arms with $m=2$ tend to wind loosely in the outer parts of the disks
and tightly at the inner radii (see Figure \ref{fig:ModelABm2}). Compared
with the cases of $m=1$, the $m=2$ arms can extend to more inner radii, thus
their contributions in the velocity-delay maps are more significant. In
addition, $\Xi$ tends to be more ``banana''-like in Model B, which is similar to
the cases with $m=1$.

It is obvious that the velocity-delay maps of the spiral arms with $m=2$ are
asymmetric and different from the velocity-delay map of a simple Keplerian disk.
The distributions of the strongest responses (bright arcs/strips in Figures
\ref{fig:vdms_ModelAm2} and \ref{fig:vdms_ModelBm2}) in the maps change along
with the LOS azimuthal angle. For example, for the velocity-delay map of
$\mu_U=2.00$ and $\tilde{\sigma}_U=0.05$ in Model A, the strongest responses
tend to be in the lower right corner if $\varphi_{\rm los}=0^{\circ}$ and
rotates to the lowest place if $\varphi_{\rm los}=90^{\circ}$. 

For Model B, the arms in the central parts also contribute strong signals
in the maps (see Figure \ref{fig:vdms_ModelBm2}). The maps look inhomogeneous
and have many sub-features. The lower parts of the maps have multiple
layers (similar to lasagna) in Case I of both Model A and B. This is
a typical feature in velocity-delay maps if there are a number of arms in the
inner radius of the $\Xi$-map.

\subsection{Velocity-resolved Lags}
\label{sec:velocity_resolved_lags}

\begin{figure*}[!ht]
    \includegraphics[width = \textwidth]{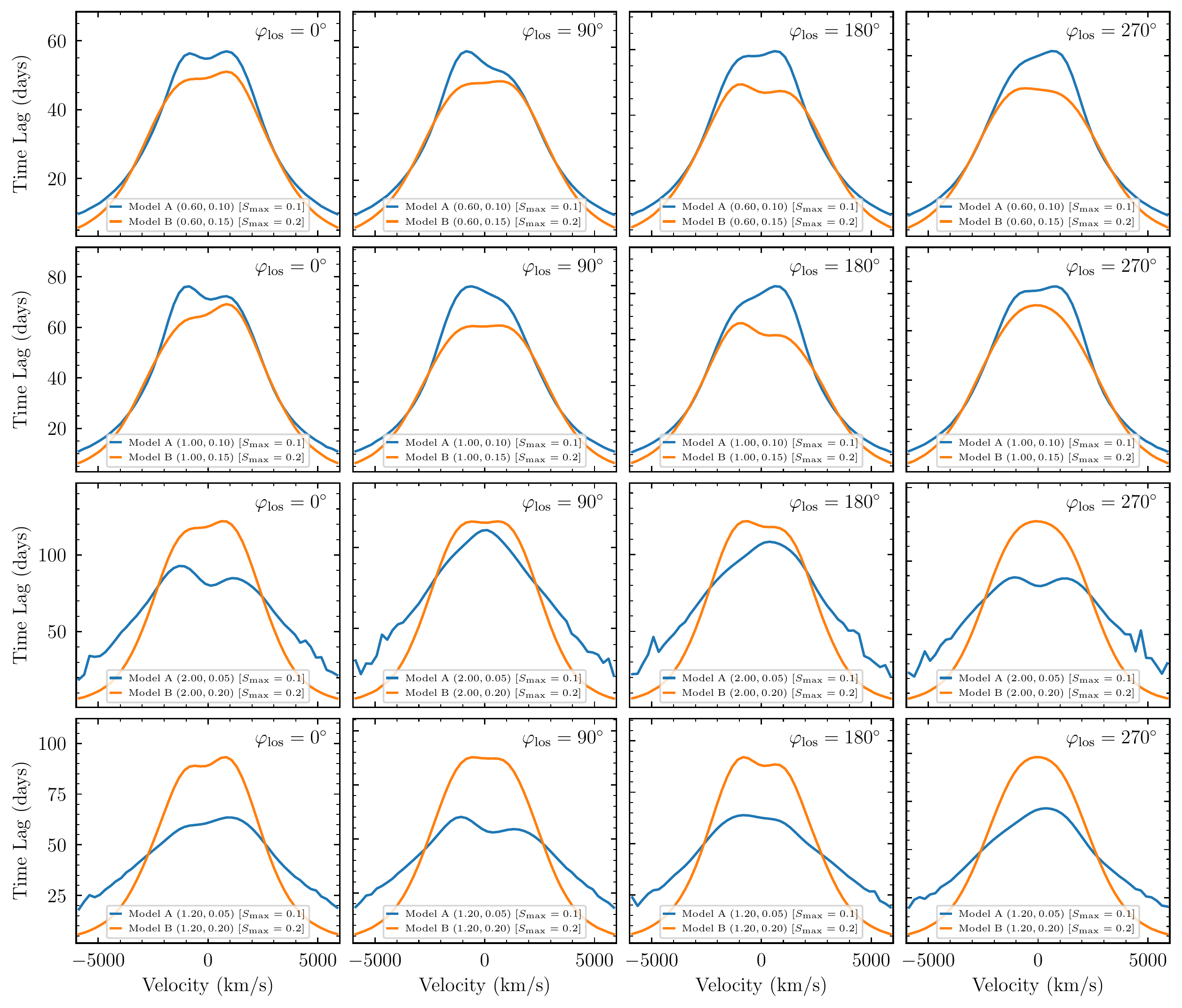}
    \caption{Velocity-resolved lags. The blue and orange lines are corresponding to Models A
    and B ($m=1$), respectively. The reprocessing coefficients and LOS azimuthal angles are marked in each panel. 
    \label{fig:velocity_resolved_lag}}
\end{figure*}

Because of the high requirement of the data quality, it is not always easy to
obtain velocity-delay maps. As a compromise, the velocity-resolved lag analysis
is also useful for the probe of BLR geometry and kinematics, and has been
applied in many RM campaigns \citep[e.g.,][]{bentz2008, bentz2009, denney2009,
denney2010, grier2013, du2016VI, du2018b, derosa2018, brotherton2020, hu2021,
lu2021, u2022, bao2022}. We present the velocity-resolved lags for Models A and
B in Cases I and II with $m=1$ by averaging the velocity-delay maps (Figures
\ref{fig:vdms_ModelA} and \ref{fig:vdms_ModelB}) along their time axes. The
results are shown in Figure \ref{fig:velocity_resolved_lag}. The blue lines are
the velocity-resolved lags of Model A, and the orange lines are the ones of
Model B. Similar to the corresponding velocity-delay maps, the velocity-resolved
lags are also asymmetric. 

Usually, the velocity-resolved lags with shorter (longer) lags in blue
velocities and longer (shorter) lags in red velocities are tend to be interpreted
as outflow (inflow). The velocity-resolved lags, which are generally disk-like
(the lags in small velocities are longer than those in high velocities) but show
asymmetries to some extend (blue or red lags are relatively shorter, called
``disk-like but with asymmetry'' hereafter), are sometimes explained by
Keplerian disks with some inflowing or outflowing velocities
\cite[e.g.,][]{derosa2018, lu2019}. However, the results shown in Figure
\ref{fig:velocity_resolved_lag} demonstrate that spiral arms can also produce
disk-like velocity-resolved lags but with some asymmetries. It implies that the
velocity-resolved lags are sometimes not enough for the diagnostic of BLR
geometry and kinematics, because they still have degeneracy. 

Here we do not plot the velocity-resolved lags for the $m=2$ arms. The arms are
$m$-fold axis-symmetric, thus their velocity-resolved lags do not have any
asymmetry.

\section{Discussions}
\label{sec:discussion}

\subsection{Emission-line Profiles: A Simple Comparison between Models and Observations}
\label{sec:compare_profiles}

\begin{figure*}[!ht]
    \includegraphics[width = \textwidth]{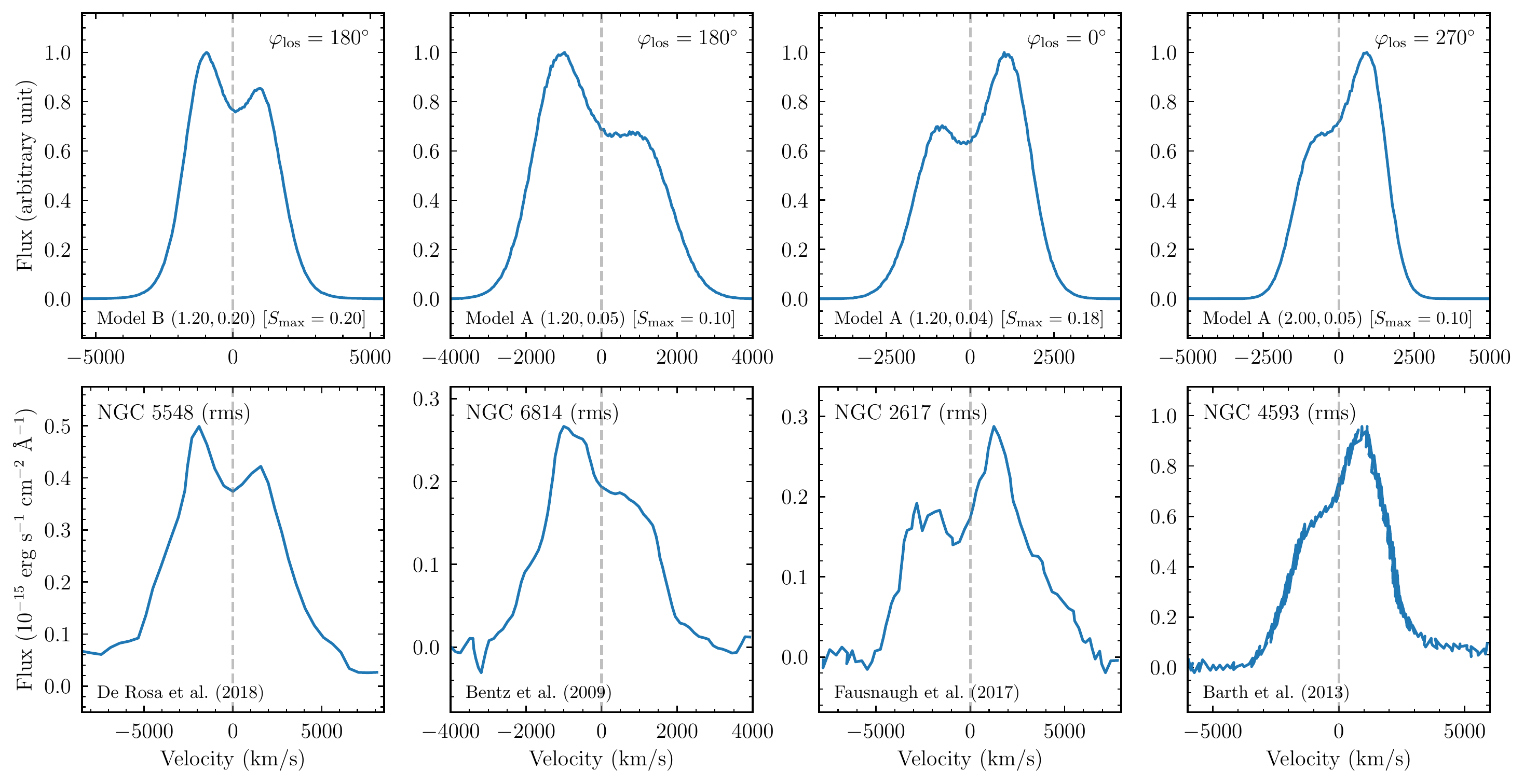}
    \caption{Some examples of the comparisons between the emission-line profiles
      generated from the models and observed in RM campaigns. The upper panels
      are the models, and the lower are the observed rms spectra scanned and
      digitized from the references marked in the lower left corners. The
      models, the parameters ($\mu_U$, $\tilde{\sigma}_U$, and $S_{\rm max}$),
      and the LOS azimuthal angles are marked in the lower left and upper right
      corners in the upper panels. The names of the objects are provided in the
      lower panels. \label{fig:profile_compare}}
\end{figure*}

In observations, the asymmetries of the emission-line profiles in single-epoch
spectra have been reported for a number of Seyfert galaxies and quasars since
1970s \citep[e.g.,][]{osterbrock1977, derobertis1985, boroson1992, marziani1996,
marziani2003, brotherton1996}. The asymmetries include, e.g., the single-peaked
profile whose peak is blueshifted or redshifted, the single-peaked profile which
has a stronger blue or red wing but a zero-velocity peak, and the double-peaked
profile with a stronger blue or red peak. Some models have been proposed to
explain the asymmetries of the emission-line profiles in AGNs.
\cite{capriotti1979} and \cite{capriotti1981} suggested that the optically-thick
clouds with inflow or outflow velocities in BLRs can produce asymmetric broad
emission lines. \cite{ferland1979} also proposed that a stronger red wing can be
explained by the self absorption of the line radiation in an expanding BLR with
optically-thick clouds. \cite{raine1981} established a disk BLR model
illuminated by the scattered radiation from the wind, which can yield slight
asymmetric line profiles. The double-peaked, asymmetric line profiles can be
explained by a relativistic Keplerian disk \citep{chen1989}.
\cite{eracleous1995} suggested that an elliptical BLR disk can interpret the
double-peaked profiles whose red peak is stronger than the blue one, which is
contrary to the prediction of a relativistic disk. More recently,
\cite{storchi-bergmann2003, storchi-bergmann2017, schimoia2012} proposed that
the spiral arms can explain the double-peaked, asymmetric line profiles and
their variations, but based on the mathematical models which presume the
analytical forms of the perturbation rather than a physical model such as in the
present paper. In addition, the asymmetries of the line profiles can also be
attributed to supermassive binary black holes \citep[e.g.,][]{shen2010, bon2012,
li2016, Ji2021}. The physical model of density waves in this paper can produce 
the double-peaked and asymmetric line profiles as those in Figures
\ref{fig:profiles_ModelA} and \ref{fig:profiles_ModelB}. 

More importantly, if the emissivity distributions of the mean and rms
spectra are different (it's always this case in observations), the line profiles
of the mean and rms spectra in the BLR spiral-arm models of the present papers
can naturally produce very different asymmetries. For instance, the mean spectrum has a
blue asymmetry but the rms spectrum has a red asymmetry, or one is generally
symmetric but the other is significantly asymmetric (see Figures
\ref{fig:profiles_ModelA} and \ref{fig:profiles_ModelB}). In observations, the
mean and rms spectra in many objects have very different line asymmetries
\citep[e.g., Mrk 202, Mrk 704, 3C 120, NGC 2617, NGC 3227, NGC 3516, NGC 4151,
NGC 4593, NGC 5548, NGC 6814,SBS 1518+693 in][]{peterson1998, bentz2009,
denney2009, grier2012, barth2013, fausnaugh2017, du2018b, derosa2018,
brotherton2020}. The BLR model with spiral arms is a very promising mechanism
that can easily explain the differences of the line profiles in the mean and rms
spectra of RM campaigns. 

Fitting the observed mean or rms line profiles with the present model is beyond
the purpose of this paper. We simply select some line profiles from our Models A
and B (without any fine-tuning), and then discover that it is easy to find some
observed rms spectra that have almost the same profiles as these models. Some simple
comparisons between the profiles of models and observations are provided in Figure
\ref{fig:profile_compare}.

The vertical radiation pressure may drive some gas flow from
the disk surface \citep[e.g.,][]{wang2012, czerny2017, elvis2017}. This
potential gas flow may contribute some velocity broadening or extra blueshift
asymmetry to the line profiles (may also influence the velocity-resolved lags
and velocity-delay maps). This effect will be considered in more details in the
future.

\subsection{Velocity-delay Map of NGC 5548 and Implications to BLR Spiral Arms}
\label{sec:NGC5548_vdm}

The high-quality velocity-delay maps of the H$\beta$ emitting region in NGC 5548
have been successfully reconstructed by the maximum entropy method in two RM
campaigns in 2014 and 2015, and are presented in \cite{horne2021} and
\cite{xiao2018b}, respectively. The two maps in 2014 and 2015 are very similar,
and both of them show traditional ``bell''-like envelopes with a bright
``elliptical rings'' which is the typical signature of a simple Keplerian disk.
However, the responses at the red velocities ($\sim2000$ km~s$^{-1}$) and long
time lags ($\sim30$ days) are relatively weaker than the other parts in both of
the two maps (\citealt{horne2021} calls it an incomplete ellipse).
\cite{xiao2018b} suggested that this weak response is due to the inhomogeneity
of the outer part of the BLR in NGC 5548. In addition,
\cite{horne2021} presents a helical ``barber-pole'' pattern in the C {\sc iv}
line of NGC 5548, which also implies the potential existence of some azimuthal
structures in the BLR.

The spiral arms stimulated from the self-gravity instabilities are probably a
physical origin of the weak response (incomplete ellipse) in the velocity-delay
map of NGC 5548. The velocity-delay map produced by Model A with
$\mu_U=0.60$, $\tilde{\sigma}_U=0.10$ (or $\mu_U=1.00$,
$\tilde{\sigma}_U=0.10$), and $\varphi_{\rm los}=90^{\circ}$ (shown in Figure
\ref{fig:vdms_ModelA}) has a similar weak response at red velocities and long
time lags (incomplete ellipse). We will carry out detailed fitting to the
velocity-delay map of NGC 5548 with the spiral-arm model in a separate paper in
future. 

\subsection{Changes of Emission-line Profiles and Velocity-resolved Lags: Arm Rotation, 
Changes of Emissivity/Responsivity, or Instabilities of Spiral Arms}
\label{sec:change_of_profile_and_vlags}

The real part of eigenvalues $\omega$ represents the rotation speed of the
spiral arms, and depend on $M_{\rm disk} / M_{\bullet}$, $\bar{Q}$, $R_{\rm
out}/R_{\rm in}$, and the inner/outer radius. We provide the values of $\omega$
in Figures \ref{fig:ModelAm1}, \ref{fig:ModelBm1}, and \ref{fig:ModelABm2}. The
timescale $2\pi/\omega$, that the arms rotate $360^{\circ}$, spans from $\sim70$
years to $\sim110$ years for the cases with $M_{\rm disk} / M_{\bullet}=0.8$ in
the present paper. However, as shown in Figures \ref{fig:profiles_ModelA},
\ref{fig:profiles_ModelB}, and \ref{fig:velocity_resolved_lag}, the
emission-line profiles and the velocity-resolved lags (or even velocity-delay
maps) can vary significantly if $\varphi_{\rm los}$ changes $90^{\circ}$. Thus,
observers will discover that the emission-line profiles and the
velocity-resolved lags (or even velocity-delay maps) change significantly in
$\sim20-30$ years if the BLR has similar parameters we adopted here
($M_{\bullet}=10^8 M_{\odot}$ and $\dotm=1.0$). Even if the parameters are
different and the spiral arms prefer different mode (see Appendix
\ref{app:eigenvalues}), the timescale can decrease further (even smaller than
$\sim10$ years). From Appendix \ref{app:eigenvalues}, the
real part of $\omega$ is generally on the order of $(G M_{\bullet} / R_{\rm
out}^3)^{1/2}$ (or larger than $(G M_{\bullet} / R_{\rm out}^3)^{1/2}$ by
factors of a few), where $(G M_{\bullet} / R_{\rm out}^3)^{1/2}$ is the
Keplerian rotation frequency at the outer radius of the disk. The rotation
timescale may be roughly $\propto (L^{3/2} / M_{\bullet})^{1/2} \propto
M_{\bullet}^{1/4} \dotm^{3/4}$. Therefore, the rotation timescale may be smaller
if the accretion rate and BH mass are smaller.

In observations, the emission-line profiles (mean or rms) and the
velocity-resolved time lags have shown significant changes between two campaigns
several to ten years apart. For instance, the line profile in the rms spectrum
of NGC 3227 was symmetric and double-peaked in 2007 \citep{denney2009}, but
became asymmetric and single-peaked (the peak is redshifted) with strong blue
wing in 2017 \citep{brotherton2020}. Its velocity-resolved lags changed from
shorter in blue and longer in red velocities to inverse from 2007 to 2017
\citep{denney2009, brotherton2020}. The velocity-resolved lags of NGC 3516
changed from longer in blue and shorter in red to inverse to some extent from
2007 to 2012 \citep{denney2009, derosa2018}, and changed back in 2018-2019
\citep{feng2021}. Considering their smaller black hole masses, the timescales of
these changes are generally consistent with the rotation timescale of the
density waves. The spiral arms in BLR is probably a very natural explanation for
such quick changes. In particular, some of the periodic variations in the line
profiles (or in the velocity-resolved lags or velocity-delay maps in future
observations) can probably be explained by the spiral arms. Future detailed
modeling will reveal the surface densities and azimuthal angles of the spiral
arms in those objects. 

Furthermore, if the continuum luminosities vary, the emissivity/responsivity
distributions may change accordingly because of the photoionization physics
(e.g., $\mu_U$, $\tilde{\sigma}_U$ may be different). In this case, the line
profiles, velocity-resolved lags, and velocity-delay maps can show significant
changes within even shorter time scales (light-traveling time scale). Therefore,
it must be crucial to monitor an object (especially the ones with large
variations, or even changing-look AGNs) repeatedly in different luminosity
states.

Finally, the instabilities of spiral arms can also be a mechanism for the short
timescales of the changes in the emission-line profiles (single-epoch, mean, or
rms) and the velocity-resolved lags. The growth rates can be comparable to the
Keplerian timescales at outer radii, especially for Model A (for Model B, the
growth timescale is longer than the Keplerian timescale by factors of a few to
ten, see Figures \ref{fig:EigenAm1}-\ref{fig:EigenABm2} in Appendix
\ref{app:eigenvalues}), which means that the timescales of the instabilities for
spiral arms can be relatively short. As mentioned above, the line profiles and
the velocity-resolved lags can change within a period as short as $\lesssim10$
years (e.g., NGC~3227, NGC~3516). In addition, the line profiles (single-epoch,
mean, or rms spectra) of some objects (e.g., Mrk~6 in \citealt{doroshenko2012}
and \citealt{du2018b}, 3C~390.3 in \citealt{sergeev2020} and \citealt{du2018b})
also showed obvious changes, but in longer timescales of $\sim20-30$ years. The
instabilities of spiral arms can also be a possible explanation for those
changes. But it should be noted that the growth timescale is still significantly
longer than the rotation timescale (see Figures
\ref{fig:EigenAm1}-\ref{fig:EigenABm2}), thus the changes caused by the
instabilities of arms may be slower than those caused by the rotation. Moreover,
the changes caused by the instabilities should be more chaotic, but the those
caused by the rotation should be ordered and probably periodic.

\subsection{Observational Tests in Future}

As shown above, directly searching the spiral-arm signatures from the
velocity-delay maps and emission-line profiles in RM campaigns is a very
promising way to identify the spiral arms in BLRs. Recently, a trend with RM
campaigns is to focus on a specific subclass of AGNs in order to investigate
their unique properties, e.g., ``Monitoring AGNs with H$\beta$ Asymmetry''
(MAHA) project targets to the AGNs with asymmetric H$\beta$ emission lines
\citep{du2018b, brotherton2020, bao2022}. We may identify some BLRs with spiral
arms from the velocity-delay maps or emission-line profiles in MAHA project in
the future. In addition, it is also promising to search candidates of spiral-arm
BLRs from some spectroscopic samples of the AGNs with asymmetry emission-line
profiles \citep[e.g.,][]{eracleous2012}. 

Furthermore, RM to some AGNs with very large flux variations may be helpful. The
velocity-delay maps of a same object at high and low states can probe different
radii of its BLR (high state for larger radius and low state for smaller
radius), and will provide a better constraints to spiral-arm pattern.

\subsection{Roles of Parameters $Q$ and $M_{\rm disk} / M_{\bullet}$}
\label{sec:roles_of_Q_and_mass_ratio}

In Section \ref{sec:spiral_arms}, we found that the spiral arms wind more
loosely if the Toomre parameter $Q$ and the mass ratio $M_{\rm disk} /
M_{\bullet}$ are larger. This phenomenon is easy to understand. The dispersion
relation of the gravitational instabilities can be expressed, in lowest
approximation, as $(\omega - m \Omega)^2 = \kappa^2 + (k a_0)^2 - 2 \pi G
\sigma_0 |k|$, where $k$ is the wave number \citep{Lin1979}. The waves are
trailing if $k<0$. The solution of the dispersion relation is $k = -k_0 [1 \pm
\sqrt{1 - Q^2 (1 - \nu^2)}]$, where $k_0 = \kappa^2 / \pi G \sigma_0 Q^2 $.
Considering that $M_{\rm disk} / M_{\bullet}$ is proportional to $\sigma_0$, the
wave number $|k|$ decreases and the wavelength increases (arms wind more
loosely) if $Q$ and $M_{\rm disk} / M_{\bullet}$ are larger.

\subsection{Linear Analysis and $\sigma_1/\sigma_0$}

As a first step, we adopted the linear analysis to describe the density wave in
disk-like BLRs and neglect the viscosity in the present paper for simplicity.
The absolute amplitude of $\sigma_1$ cannot be directly deduced from Eqn
(\ref{eq:integro-differential}) and is freely scalable (the solution of Eqn
\ref{eq:matrix_5N} can be $S_l^*$ or $C S_l^*$ with an arbitrary constant $C$).
In more realistic calculations, the dissipation processes such as shocks,
nonlinear growth of perturbations, or viscous stress should be taken into
account. On one hand, the dissipation can lead to a deposit of the angular
momentum carried by density wave to the disk, which may also induce changes in
the surface density of the disk. On the other hand, the absolute amplitude of
$\sigma_1$ may be determined if the growth of perturbation becomes saturated by
the dissipation processes \citep[e.g.,][]{laughlin1996, laughlin1997}. These
effects are not included in current equations of motion (Eqn \ref{eq:motion_1}
and \ref{eq:motion_2}) and the normal mode matrix equation (Eqn
\ref{eq:matrix_5N}), and will be considered further in future.

\subsection{Accretion Driven by Spiral Arms}

The dimensionless accretion rate $\dot{\mathscr{M}}$ is only used to determine
the continuum luminosity and further the inner and outer radii, as well as the
appropriate reference parameters for $\Xi$ in Table
\ref{tab:reprocessing_coefficient}. We mainly focus the spiral arms in BLRs
which typically span from $10^3 R_{\rm g}$ to $10^5 R_{\rm g}$. The
UV/optical continuum luminosity comes from the more inner accretion disk
($\lesssim 10^3 R_{\rm g}$), which could be in Shakura \& Sunyaev regime
\citep{shakura1973}. Discussing the angular momentum transfer in details is
beyond the scope of this paper. However, we can roughly evaluate if the
accretion rate driven by the spiral structures in these regions is enough for
the accretion in the inner disk. 

In a viscous thin disk with quasi-Keplerian rotation, the radial velocity
induced by a viscosity $\nu_{\rm vis}$ \citep{lynden-bell1974} can be
expressed as
\begin{equation}
    u = \left[\sigma_0 R \frac{\partial \Omega R^2}{\partial R}\right]^{-1} 
        \frac{\partial}{\partial R}\left[\sigma_0 \nu_{\rm vis} R^3 \frac{\partial \Omega}{\partial R}\right]
    \sim \alpha a_0 \frac{H}{R},
\end{equation}
where $\nu_{\rm vis}=\alpha a_0 H$ is an effective ``alpha''-type viscosity,
$\alpha$ is viscosity parameter, and $H$ is the thickness of the disk. The mass
accretion rate can be obtained with $\dot{M_{\bullet}}=2 \pi R u \sigma_0$. The
global spiral arms may redistribute the disk material and be described in terms
of a diffusive process with an effective viscosity $\alpha_{\rm eff}$
\citep{laughlin1996}. $\alpha_{\rm eff}$ is on the order of 0.01 or so
(especially in nonlinear regime, e.g., \citealt{laughlin1994, laughlin1996,
lodato2005}). We have checked that, with such a $\alpha_{\rm eff}$, the disk
properties assumed in the present paper ($\sigma_0$, $a_0$, $H$, and $\Omega$)
can very easily support the accretion with $\dot{\mathscr{M}}\sim$ 1.

\subsection{Vertical Structures and Possible Influences}

Given the sound speed $a_0$ and rotation curve $\Omega$, the thickness of the
disk is $H/R \sim R^{1/8}$ and $H/R \sim R^{1/4}$ for Model A and B,
respectively. It means that the geometry of the disk is ``bowl-shaped''
(concave, see \citealt{starkey2022}). Such geometry can enable the disk to be
illuminated by the ionizing photons from the inner region.

With surface density ($\sigma_1$) variations, the disk thickness is also likely
to modulate. The wave crest of the arm may be more strongly irradiated by the
ionizing photons because it protrudes from the disk surface. On the contrary,
the wave trough may be more weakly irradiated. Therefore, the asymmetries of the
line profiles and velocity-resolved lags, and the sub-features in the
velocity-delay maps may be more stronger. A sophisticated treatment of the
vertical structures and the corresponding influences to the observation are
needed in the future.

\subsection{Boundary Conditions}

In this paper, we adopted the same boundary conditions as in \cite{adams1989}
for simplicity. \cite{noh1991} and \cite{chen2021} investigated the influence of
boundary conditions on the pitch angles, pattern speeds, and growth rates of
spiral arms in protoplanetary disks. They tried reflecting and transmitting
boundaries besides the boundary conditions of \cite{adams1989}, and found that
the boundary conditions mainly influence the growth rates but have little effect
on the pitch angles and pattern speeds of the arms (the differences are
$\lesssim10\%$ for different boundary conditions). Their works indicate that
adopting the boundary conditions of \cite{adams1989} is enough for exhibiting
the general reverberation properties of the BLR arms in observations. In the
future, the boundary conditions may be revised by comparing the models with the
real observations.

\section{Summary}
\label{sec:summary}

In recent years, there are growing evidences that some of BLRs are inhomogeneous
and have substructures. The radii of BLRs measured by RM are consistent with the
self-gravitating regions of accretion disks, which implies that the spiral arms
excited by the gravitational instabilities may exist in, at least, the disk-like
BLRs. In this paper, we calculate the surface densities of the spiral arms in
BLRs, for two typical configurations (called Models A and B) with different
parameters, by using the density wave theory. We find that more massive disks
(larger disk-to-SMBH mass ratios) with larger Toomre parameters tend to have more
loosely wound arms (more significant in observations). In comparison with Model
A, the spiral arms of Model B are more ``banana''-like. 

We present the emission-line profiles, velocity-delay maps, and
velocity-resolved lags for the cases of loosely wound spiral arms (in more
massive BLR disks). For $m=1$ spiral arms, the emission-line profiles,
velocity-resolved lags have significant asymmetries, and the velocity-delay maps
are asymmetric and have complex substructures (bright arcs/strips). For $m=2$
spiral arms, the emission-line profiles and velocity-resolved lags are
symmetric, on the contrary, the velocity-delay maps are asymmetric and show complex
substructures. The spiral arms in BLRs can easily explain some phenomena in
observations:

\begin{itemize}
    \item For a same object, the mean and rms spectra in RM observations can
       have very different asymmetries. The rms spectra always have different
       widths compared to the mean spectra in RM campaigns, which implies that
       the emissivities/responsivities of the invariable and variable parts in
       BLRs are different. Considering the different
       emissivities/responsivities, the calculations in the present paper show
       that the spiral arms in BLRs can naturally produce  
       differently-asymmetric line profiles in the mean and rms spectra of a
       same object without any further special assumptions.

    \item Our models can generate emission-line profiles almost the same as
      the observations (rms spectra). 

    \item The spiral arms in the disk-like BLRs can produce complex features
      such as bright arcs/strips, and are probably a physical origin for the
      relatively-weak response region (incomplete ellipse) in the velocity-delay
      map of NGC 5548.

    \item The timescale that the spiral arms rotate $\varphi_{\rm los} \sim
      90^{\circ}$ (which can significantly changes the line profiles or
      velocity-resolved lags) can be as short as $\lesssim 10$ years. The
      rotation of the spiral arms can explain the quick changes of the
      asymmetries in the emission-line profiles, the velocity-resolved lags, or
      even velocity-delay maps between RM campaigns several to ten years apart.
      Futhermore, some of the periodic variations in the line profiles (or in
      the velocity-resolved lags or velocity-delay maps in future observations)
      can probably be explained by the rotation of the BLR spiral arms.

    \item The line profiles, velocity-resolved lags, and velocity-delay
    maps can show significant changes within short time scales (light-traveling
    time scale) if the continuum vary significantly.
\end{itemize}

Sophisticated fitting to the observations by the spiral-arm models will reveal
the detailed geometry and kinematics of BLRs in the future.

\begin{acknowledgements}
We thank the anonymous referee for the useful comments that
improved the manuscript. We acknowledge the support by National Key R\&D
Program of China (grants 2021YFA1600404, 2016YFA0400701), the support by the
National Science Foundation of China through grants {NSFC-12022301, -11991051,
-11991054, -11873048, -11833008}, by Grant No. QYZDJ-SSW-SLH007 from the Key
Research Program of Frontier Sciences, Chinese Academy of Sciences (CAS), by the
Strategic Priority Research Program of CAS grant No.XDB23010400, and by the
International Partnership Program of CAS, Grant No.113111KYSB20200014.
\end{acknowledgements}

\begin{appendix}

    \section{Eigenvalues}
    \label{app:eigenvalues}
    
    \subsection{Eigenvalues of Models A and B with $m=1$}
    
    \begin{figure*}[!h]
        \centering
        \includegraphics[width = 0.95\textwidth]{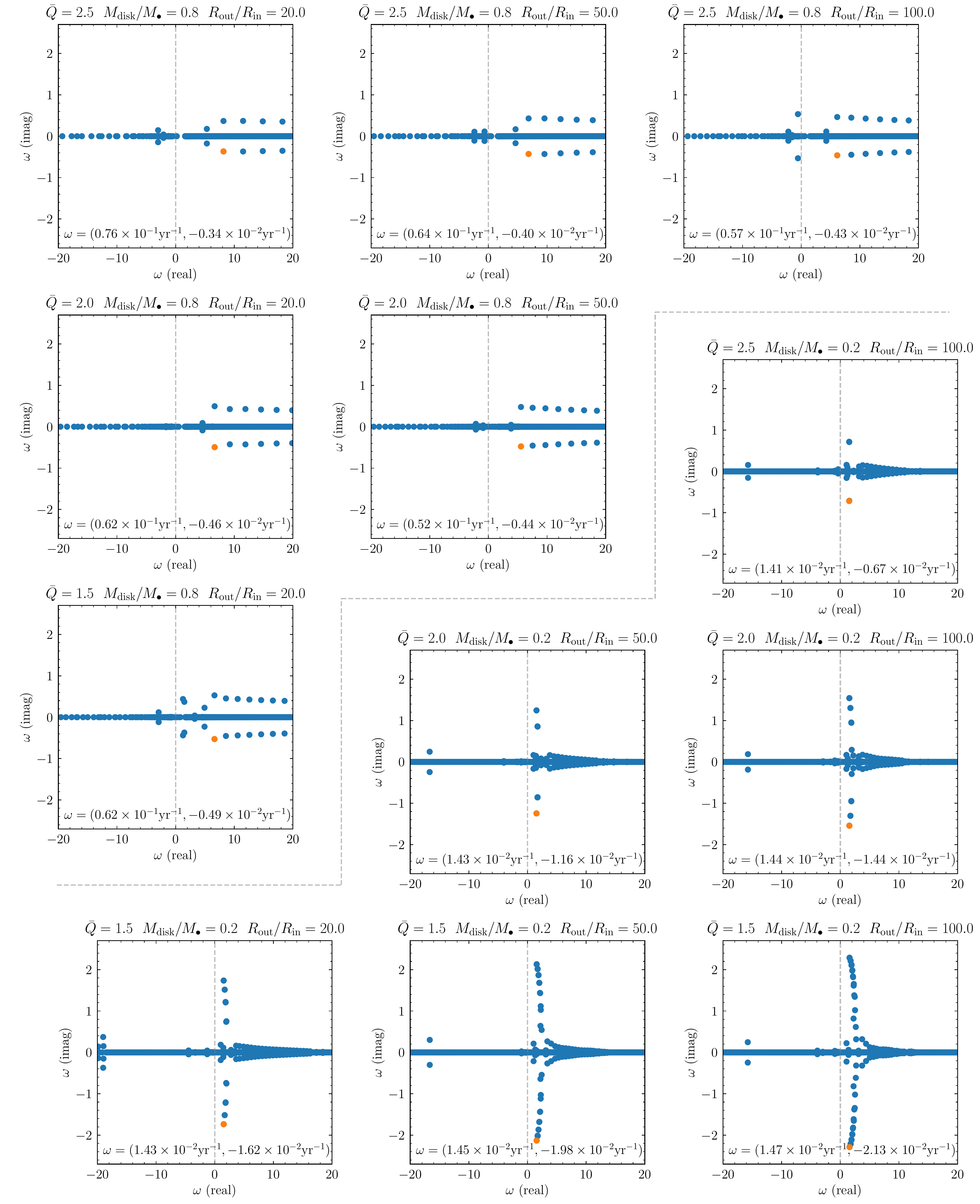}
        \caption{Eigenvalues of the spiral arms ($m=1$) for Model A. The real and
        imaginary parts of $\omega$ are both in units of $(G M_{\bullet} / R_{\rm
        out}^3)^{1/2}$. The 6 panels in upper left corner are the eigenvalues for
        more massive disks ($M_{\rm disk}/M_{\bullet}=0.8$), and the 6 panels in
        lower right corner are those for less massive disks ($M_{\rm
        disk}/M_{\bullet}=0.2$). The values of $\bar{Q}$, $M_{\rm
        disk}/M_{\bullet}$, and $R_{\rm out}/R_{\rm in}$ are marked on the top of
        each panels. The eigenvalue adopted in the present paper is marked in orange
        in each panel, and its value is also provided in the same panel.
        \label{fig:EigenAm1}}
    \end{figure*}
    
    \begin{figure*}[!h]
        \centering
        \includegraphics[width = 0.95\textwidth]{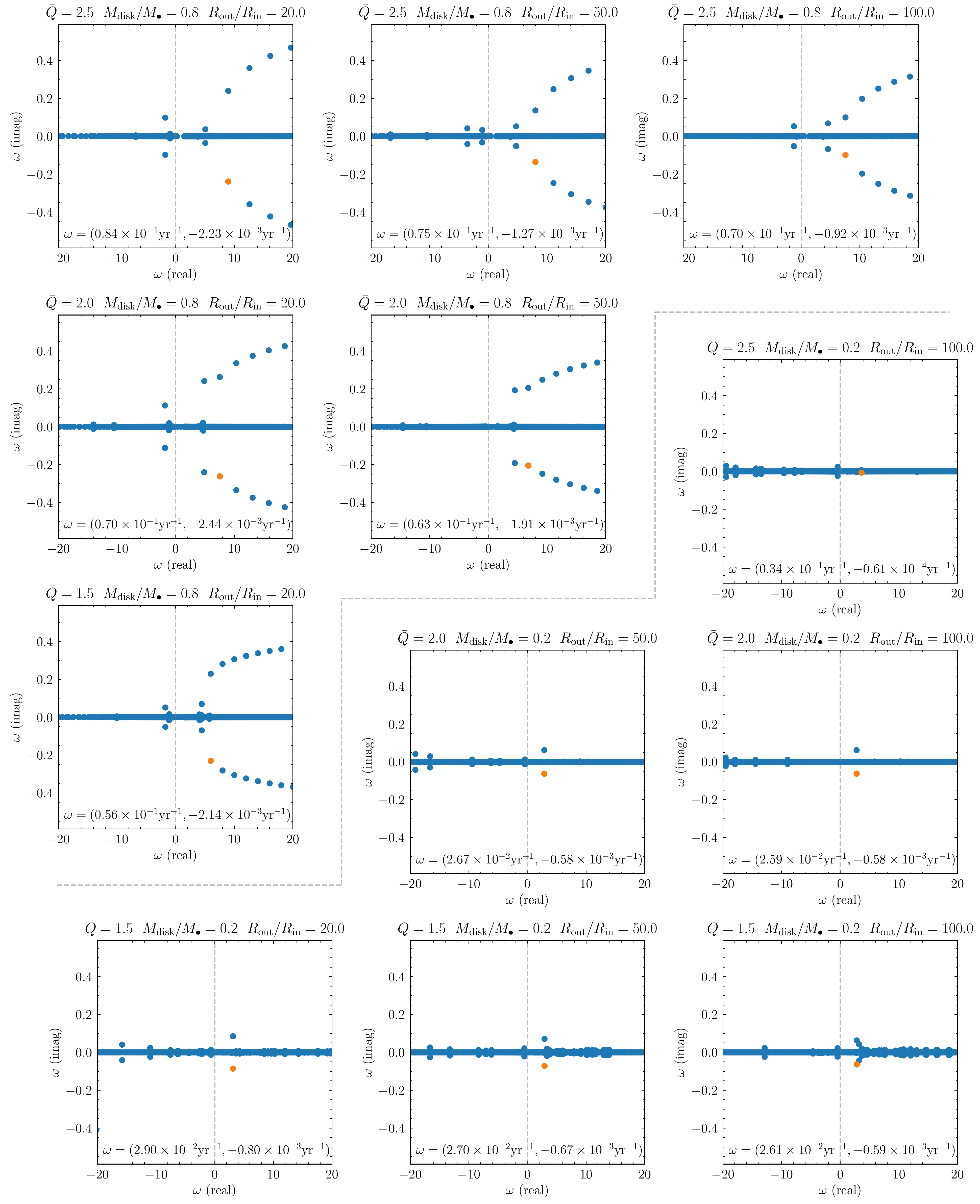}
        \caption{Eigenvalues of the spiral arms ($m=1$) for Model B. Similar to
        Figure \ref{fig:EigenAm1}, the real and imaginary parts of $\omega$ are both
        in units of $(G M_{\bullet} / R_{\rm out}^3)^{1/2}$. The 6 panels in upper
        left corner are the eigenvalues for more massive disks ($M_{\rm
        disk}/M_{\bullet}=0.8$), and the 6 panels in lower right corner are those
        for less massive disks ($M_{\rm disk}/M_{\bullet}=0.2$). The values of
        $\bar{Q}$, $M_{\rm disk}/M_{\bullet}$, and $R_{\rm out}/R_{\rm in}$ are
        marked on the top of each panels. The eigenvalue adopted in the present
        paper is marked in orange in each panel, and its value is also provided in
        the same panel.  \label{fig:EigenBm1}}
    \end{figure*}
    
    We present here the eigenvalues of Models A and B with $m=1$ in Figures
    \ref{fig:EigenAm1} and \ref{fig:EigenBm1}. For each set of parameters, there are
    more than one eigenvalues and solutions (modes). The excitation and the
    evolution of the modes (which mode will finally dominate) have not been fully
    understood \citep{bertin1989}. We adopt the mode with the lowest order and
    significant growth rate because they will be the most global and can grow in a
    relatively rapid rate \citep[e.g.,][]{adams1989, chen2021}. For consistency, we
    select the same mode for the same disk-to-SMBH mass ratio ($M_{\rm
    disk}/M_{\bullet}$) in order to demonstrate how the spiral-arm pattern evolves
    if the other parameters change, but have checked that it does not change the
    main conclusions in the present paper if we adopt the other nearby mode. For instance,
    the lowest mode become significant if $M_{\rm disk} / M_{\bullet}=0.8$,
    $\bar{Q}=1.5$, and $R_{\rm out}/R_{\rm in}=20$ in Model A (see Figure
    \ref{fig:EigenAm1}). We select the mode with the relatively higher growth rate
    rather than the lowest mode (see Figure \ref{fig:EigenAm1}). For Model B, the
    lowest modes do not have the highest growth rate. We adopt the lowest mode with
    relatively significant growth rate (also keep selecting the same mode for the
    same disk-to-SMBH mass ratio). 
    
    \subsection{Eigenvalues of Models A and B with $m=2$}
    
    \begin{figure*}[!h]
        \centering
        \includegraphics[width = \textwidth]{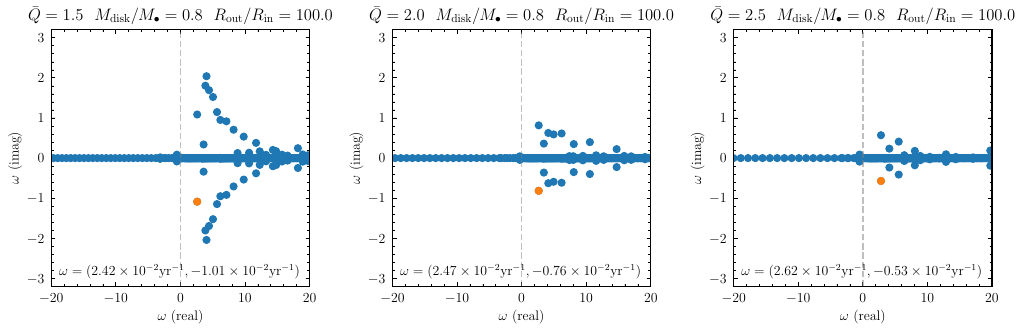}
        \includegraphics[width = \textwidth]{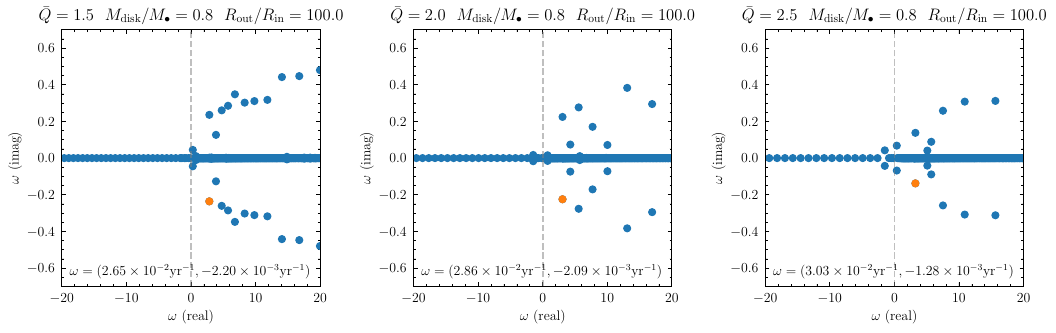}
        \caption{Eigenvalues of the spiral arms ($m=2$) for Models A and B. The real
        and imaginary parts of $\omega$ are both in units of $(G M_{\bullet} /
        R_{\rm out}^3)^{1/2}$. The values of $\bar{Q}$, $M_{\rm disk}/M_{\bullet}$,
        and $R_{\rm out}/R_{\rm in}$ are marked on the top of each panels. The
        eigenvalue adopted in the present paper is marked in orange in each panel,
        and its value is also provided in the same panel. \label{fig:EigenABm2}}
    \end{figure*}
    
    The eigenvalues of Models A and B with $m=2$ are shown in Figure
    \ref{fig:EigenABm2}. Similarly, we tend to select the lowest mode with
    relatively significant growth rate.

    \clearpage
    
    \section{Matrixes in Numerical Method}
    \label{app:matrix}
    
    The details of the numerical method for solving the governing
    integro-differential equation (Eqn \ref{eq:integro-differential}) are provided
    in \cite{adams1989}. However, the coefficients in the matrixes
    $\mathscr{W}^{(0)}$, $\mathscr{W}^{(1)}$, $\mathscr{W}^{(2)}$,
    $\mathscr{W}^{(3)}$, $\mathscr{W}^{(4)}$, $\mathscr{W}^{(5)}$ regrouped from
    $\mathscr{W}_{ik}$ are not provided (see Eqn B12 in \citealt{adams1989}). In
    this section, we demonstrate the coefficients of these matrixes. The same
    nomenclatures as \cite{adams1989} are adopted here, except the radius, the mass
    of central object, and the mass of disk are $R$, $M_{\bullet}$, $M_{\rm disk}$
    in the present paper and $r$, $M_{*}$, $M_{\rm D}$ in \cite{adams1989},
    respectively. The matrixes $\mathscr{W}^{(0)}$, $\mathscr{W}^{(1)}$,
    $\mathscr{W}^{(2)}$, $\mathscr{W}^{(3)}$, $\mathscr{W}^{(4)}$,
    $\mathscr{W}^{(5)}$ can be expressed as
    \begin{equation}
        \label{eqn:W}
        \begin{aligned}
        \mathscr{W}^{(0)} &= \mathscr{F}^{(0)}, \\
        \mathscr{W}^{(1)} &= \mathscr{F}^{(1)}, \\
        \mathscr{W}^{(2)} &= \mathscr{F}^{(2)} + \mathscr{G}^{(0)}, \\
        \mathscr{W}^{(3)} &= \mathscr{F}^{(3)} + \mathscr{G}^{(1)}, \\
        \mathscr{W}^{(4)} &= - C_2^{(4)} \frac{\kappa^2 R}{2 \pi G \sigma_0} \delta_{ik} + \mathscr{G}^{(2)}, \\
        \mathscr{W}^{(5)} &= - C_2^{(5)} \frac{\kappa^2 R}{2 \pi G \sigma_0} \delta_{ik} + \mathscr{G}^{(3)},
        \end{aligned}
    \end{equation}
    where
    \begin{equation}
        \label{eqn:Fn}
        \begin{split}
        \mathscr{F}^{(n)} = \Bigl\{ C_n \mathscr{D}_{ij}^{(2)} + \left[C_A^{(n)} + C_n \left[2(1-p) - 1\right]\right] \mathscr{D}_{ij}^{(1)} \\
        + \left[C_B^{(n)} + C_A^{(n)} (1-p) + C_n p (p-1)\right] \delta_{ij} \Bigr\} \mathscr{I}_{jk} \\
        + \frac{1}{\Sigma R} \Bigl\{ C_n \mathscr{D}_{ik}^{(2)} + \left[ C_A^{(n)} - C_n (2q + 1) \right] \mathscr{D}_{ik}^{(1)} \\
        + \left[ C_B^{(n)} - C_A^{(n)} q + C_n q (q+1) \right] \delta_{ik} \Bigr\} \\
        - C_2^{(n)} \frac{\kappa^2 R}{2 \pi G \sigma_0} \delta_{ik},  \ \ \ \ n = 0, 1, 2, 3, 
        \end{split}
    \end{equation}
    and
    \begin{equation}
        \label{eqn:Gn}
        \mathscr{G}^{(n)} = \delta_{1m} \left( C_A^{(n)} + C_B^{(n)} \right) \frac{R^3}{2 G (M_{\bullet} + M_{\rm disk})} \mathscr{J}_{ik}, \ \ \ \ n = 0, 1, 2, 3.
    \end{equation}
    $\mathscr{D}_{ij}^{(1)}$ and $\mathscr{D}_{ij}^{(2)}$ are
    the first- and second-order derivatives expressed in matrix form, see Eqn
    (B4a) and (B4b) in \cite{adams1989}. $\mathscr{I}_{jk}$ and
    $\mathscr{J}_{ik}$ are two matrixes that perform integrals, see also the
    appendix in \cite{adams1989}. The coefficients in Eqn (\ref{eqn:W}),
    (\ref{eqn:Fn}), and (\ref{eqn:Gn}) are 
    \begin{equation}
        \begin{split}
        C_0 &= \frac{m^3 \Omega^3 - m \Omega \kappa^2}{\kappa^3}, \\
        C_1 &= \frac{\kappa^2 - 3 m^2 \Omega^2}{\kappa^3}, \\
        C_2 &= \frac{3 m \Omega}{\kappa^3}, \\
        C_3 &= -\frac{1}{\kappa^3}, 
        \end{split}
    \end{equation}
    \begin{equation}
        \begin{aligned}
        C_A^{(0)} &= C_0 \mathscr{D}_{ij}^{(1)} \left[\log(\sigma_0 R)\right]_{j} + \frac{m \Omega}{\kappa^3} \mathscr{D}_{ij}^{(1)} (\kappa^2)_{j} - \frac{2 m^3 \Omega^2}{\kappa^3} \mathscr{D}_{ij}^{(1)} \Omega_{j}, \\
        C_A^{(1)} &= C_1 \mathscr{D}_{ij}^{(1)} \left[\log(\sigma_0 R)\right]_{j} - \frac{1}{\kappa^3} \mathscr{D}_{ij}^{(1)} (\kappa^2)_{j} + \frac{4 m^2 \Omega}{\kappa^3} \mathscr{D}_{ij}^{(1)} \Omega_{j}, \\
        C_A^{(2)} &= C_2 \mathscr{D}_{ij}^{(1)} \left[\log(\sigma_0 R)\right]_{j} - \frac{2m}{\kappa^3} \mathscr{D}_{ij}^{(1)} \Omega_{j}, \\
        C_A^{(3)} &= C_3 \mathscr{D}_{ij}^{(1)} \left[\log(\sigma_0 R)\right]_{j}, 
        \end{aligned}
    \end{equation}
    \begin{equation}
        \begin{aligned}
            C_B^{(0)} &= -m^2 C_0 - \frac{4 m^3 \Omega^2}{\kappa^3} \mathscr{D}_{ij}^{(1)} \Omega_j + \frac{4 m^3 \Omega^3}{\kappa^4} \mathscr{D}_{ij}^{(1)} \kappa_j \\
                      &\quad + \frac{2 m \Omega}{\kappa^3} \left( \kappa^2 - m^2 \Omega^2 \right) \mathscr{D}_{ij}^{(1)} \left(\log \frac{\kappa^2}{\Omega \sigma_0}\right)_j, \\
            C_B^{(1)} &= -m^2 C_1 + \frac{4 m^2 \Omega}{\kappa^3} \mathscr{D}_{ij}^{(1)} \Omega_j - \frac{8 m^2 \Omega^2}{\kappa^4} \mathscr{D}_{ij}^{(1)} \kappa_j \\
                      &\quad + \frac{4 m^2 \Omega^2}{\kappa^3} \mathscr{D}_{ij}^{(1)} \left(\log \frac{\kappa^2}{\Omega \sigma_0}\right)_j, \\
            C_B^{(2)} &= -m^2 C_2 + \frac{4 m \Omega}{\kappa^4} \mathscr{D}_{ij}^{(1)} \kappa_j - \frac{2 m \Omega}{\kappa^3} \mathscr{D}_{ij}^{(1)} \left(\log \frac{\kappa^2}{\Omega \sigma_0}\right)_j, \\
            C_B^{(3)} &= -m^2 C_3,  
        \end{aligned}
    \end{equation}
    and
    \begin{equation}
        \begin{aligned}
        C_2^{(0)} &= -\frac{m \Omega}{\kappa} + \frac{2 m^3 \Omega^3}{\kappa^3} - \frac{m^5 \Omega^5}{\kappa^5}, \\
        C_2^{(1)} &= \frac{1}{\kappa} - \frac{6 m^2 \Omega^2}{\kappa^3} + \frac{5 m^4 \Omega^4}{\kappa^5}, \\
        C_2^{(2)} &= \frac{6 m \Omega}{\kappa^3} - \frac{10 m^3 \Omega^3}{\kappa^5}, \\
        C_2^{(3)} &= -\frac{2}{\kappa^3} + \frac{10 m^2 \Omega^2}{\kappa^5}, \\
        C_2^{(4)} &= -\frac{5 m \Omega}{\kappa^5}, \\
        C_2^{(5)} &= \frac{1}{\kappa^5}.
        \end{aligned}
    \end{equation}
    
    The first and last rows of the matrixes
    $\mathscr{W}^{(0)}$ to $\mathscr{W}^{(5)}$ are determined by the boundary conditions, and 
    only have the terms of 0 to 3 orders. The inner boundary conditions are
    \begin{equation}
        \begin{aligned}
            \mathscr{W}^{(0)}_{1k} &= \left[ -m \Omega \mathscr{D}_{1j}^{(1)} + (p-3) m \Omega \delta_{1j} \right] \mathscr{I}_{jk} \\
                                   &\quad + \frac{1}{\Sigma R} \left[ -m \Omega \mathscr{D}_{1k}^{(1)} + (q-2) m \Omega \delta_{1k} \right], \\
            \mathscr{W}^{(1)}_{1k} &= \left[ \mathscr{D}_{1j}^{(1)} + (1-p) \delta_{1j} \right] \mathscr{I}_{jk} + \frac{1}{\Sigma R} \left( \mathscr{D}_{1k}^{(1)} - q \delta_{1k} \right), \\
            \mathscr{W}^{(2)}_{1k} &= -\frac{3 \delta_{1m} \Omega R^3}{2 G (M_{\bullet} + M_{\rm disk})} \mathscr{J}_{1k}, \\
            \mathscr{W}^{(3)}_{1k} &= \frac{\delta_{1m} R^3}{2 G (M_{\bullet} + M_{\rm disk})} \mathscr{J}_{1k}.
        \end{aligned}
    \end{equation}
    The outer boundary conditions are
    \begin{equation}
        \begin{aligned}
            \mathscr{W}^{(0)}_{Nk} &= \left[ -m \Omega \mathscr{D}_{Nj}^{(1)} + (p-3) m \Omega \delta_{Nj} \right] \mathscr{I}_{jk} \\
                                   &\quad + \frac{1}{\Sigma R} \left[ -m \Omega \mathscr{D}_{Nk}^{(1)} + (q-2) m \Omega \delta_{Nk} \right] + C_0 \frac{\kappa^3 R}{2 \pi G \sigma_0 p} \delta_{Nk}, \\
            \mathscr{W}^{(1)}_{Nk} &= \left[ \mathscr{D}_{Nj}^{(1)} + (1-p) \delta_{Nj} \right] \mathscr{I}_{jk} + \frac{1}{\Sigma R} \left( \mathscr{D}_{Nk}^{(1)} - q \delta_{Nk} \right) + C_1 \frac{\kappa^3 R}{2 \pi G \sigma_0 p} \delta_{Nk}, \\
            \mathscr{W}^{(2)}_{Nk} &= C_2 \frac{\kappa^3 R}{2 \pi G \sigma_0 p} \delta_{Nk} - \frac{3 \delta_{1m} \Omega R^3}{2 G (M_{\bullet} + M_{\rm disk})} \mathscr{J}_{Nk}, \\
            \mathscr{W}^{(3)}_{Nk} &= C_3 \frac{\kappa^3 R}{2 \pi G \sigma_0 p} \delta_{Nk} + \frac{\delta_{1m} R^3}{2 G (M_{\bullet} + M_{\rm disk})} \mathscr{J}_{Nk}. 
        \end{aligned}
    \end{equation}
    
\end{appendix}

\end{document}